\documentclass[apj]{emulateapj}
\usepackage{natbib}
\usepackage{physics}
\usepackage{hyperref}
\hypersetup{
    colorlinks=true,
    linkcolor=blue,
    citecolor=blue,
    filecolor=blue,
    urlcolor=blue
}
\usepackage{amsmath,amssymb}
\usepackage{mathptmx}
\usepackage{mathtools}
\usepackage{euscript}
\DeclareMathAlphabet{\mathpzc}{T1}{pzc}{m}{it}
\DeclareSymbolFont{matha}{OML}{txmi1}{m}{it}% txfonts
\DeclareMathSymbol{\varv}{\mathord}{matha}{118}
\usepackage{graphicx}
\usepackage{color}

\newcommand\ddfrac[2]{\frac{\displaystyle #1}{\displaystyle #2}}
\newcommand\ho{\ifmmode {\rm HI} \else H{\small I} \fi}
\newcommand\hh{\ifmmode {\rm H_2} \else H$_2$ \fi}
\def\no{\ifmmode {N_{\rm HI}} \else $N_{\rm HI}$ \fi}
\def\nt{\ifmmode {N_{\rm H_2}} \else $N_{\rm HI}$ \fi}
\def\msun{\ifmmode {\rm M_{\odot}}\else $\rm M_{\odot}$\fi}
\def\mpc{\ifmmode {\rm M_{\odot} \ pc^{-2}} \else $\rm M_{\odot} \ pc^{-2}$ \fi}
\def\tra{\ifmmode  \text{HI-to-H}_2\else H{\small I}-to-H$_2$ \fi}
\def\aG{\ifmmode {\alpha G}\else $\alpha G$ \fi}
\def\iuv{\ifmmode {I_{\rm UV}}\else $I_{\rm UV}$ \fi}

\def\sg{\ifmmode \sigma_{g} \else $\sigma_{g}$ \fi}
\def\st{\ifmmode \widetilde{\sigma}_g \else $\widetilde{\sigma}_g$ \fi}
\newcommand\hd{\ifmmode \textrm{HI-dust} \else H{\small I}-dust \fi}
\DeclareMathAlphabet{\pazocal}{OMS}{zplm}{m}{n}
\newcommand\ms{\ifmmode \pazocal{M}_s \else $\pazocal{M}_s$ \fi}
\newcommand\Nm{\ifmmode  N_{\rm M}  \else $N_{\rm M}$ \fi}

\defcitealias{Sternberg2014}{S14}
\defcitealias{Bialy2016}{BS16}
\defcitealias{VazquezSemadeni2001}{VG01}

\begin{document}

\title{The \tra transition in a Turbulent Medium}
\author{Shmuel Bialy\altaffilmark{1}, Blakesley Burkhart\altaffilmark{2}, \& Amiel Sternberg\altaffilmark{1}}
\altaffiltext{1}{Raymond and Beverly Sackler School of Physics \& Astronomy, Tel Aviv University, Ramat Aviv 69978, Israel}
\altaffiltext{2}{Harvard-Smithsonian Center for Astrophysics, 60 Garden St., Cambridge, MA 0213}
\email{$^\star$shmuelbi@mail.tau.ac.il}
\slugcomment{Submitted to ApJ, 20 March  2017}

\shorttitle{\tra in a Turbulent Medium}
\shortauthors{Bialy, Burkhart \& Sternberg}

\begin{abstract}
%The transition of atomic to molecular hydrogen is fundamental for understanding star formation and galaxy evolution. This is because star formation primarily occurs in turbulent cold and dense molecular clouds and not in the atomic medium. 
%However analytic models of the \tra transition have thus far only considered uniform density distributions. 
We study the effect of density fluctuations induced by turbulence on the $\ho/\hh$ structure in photodissociation regions (PDRs) both analytically and numerically.
We perform magnetohydrodynamic numerical simulations for both subsonic and supersonic turbulent gas, and chemical $\ho/\hh$ balance calculations. We derive atomic-to-molecular density profiles and the \ho column density probability density function (PDF) assuming chemical equilibrium.
We find that while the $\ho/\hh$ density profiles are strongly perturbed in turbulent gas, the mean \ho column density is well approximated by the uniform-density analytic formula of \citet{Sternberg2014}.
The PDF width depends on (a) the radiation intensity to mean density ratio, (b) the sonic Mach number and (c) the turbulence decorrelation scale, or driving scale.
We derive an analytic model for the \ho PDF and demonstrate how our model, combined with 21 cm observations, can be used to constrain the Mach number and driving scale of turbulent gas. 
As an example, we apply our model to 
 observations of \ho in the Perseus molecular cloud. 
We show that a narrow observed \ho PDF may imply small scale decorrelation, pointing to the potential importance of subcloud-scale turbulence driving.
% using emission lines we can not make the above statement - it does not agree with a number of independent observations (spin temperature measurements and spectra
\end{abstract}

\keywords{galaxies: star formation -- photon-dominated region (PDR) -- magnetohydrodynamics: MHD}
\section{Introduction}
\label{sec: into}

%star formation and the HI-H2 transition
Giant molecular clouds serve as the nurseries for new stars in our Galaxy and in external galaxies \citep{McKee2007}.
On global scales, observations of CO and dust show that the star-formation rate (SFR) surface density ($\Sigma_{\rm SFR}$) correlates with the H$_2$ mass surface density ($\Sigma_{\rm H_2}$), following an almost linear trend \citep{Bigiel2008, Genzel2010,Schruba2011,Tacconi2013,Azeez2016}.
The presence of H$_2$ molecules is a basic ingredient for the formation of other heavy molecules such as CO, OH and H$_2$O that serve as efficient coolants of cold gas \citep[e.g.,][]{Herbst1973, Sternberg1995, Tielens2013, vanDishoeck2013a, Bialy2015MNRAS}.
% However, it is unclear if star-formation is first triggered by gas cooling via the process of H$_2$ formation or if the gas is already self-gravitating before efficient cooling takes place. 
% Simulations and observations show that the two processes most likely happen hand-by-hand in similar physical environments of high density, UV shielded gas \citep{McKee2010} 
The study of far-ultraviolet (UV) shielding and the subsequent \tra conversion is of fundamental importance for star-formation and molecule formation in the interstellar medium (ISM).

%short review of the analytic theories and their assumptions.  

 The \tra transition in the interstellar medium of galaxies has been investigated by numerous authors over the last several decades, 
through analytic and numerical modeling
\citep[e.g.,][]{Federman1979,VanDishoeck1986, Sternberg1988, Kaufman1999, Goldsmith2007, Gnedin2014, Liszt2015}, 
as well as via hydrodynamics simulations \citep[e.g.,][]{Robertson2008, Gnedin2009, Glover2010, Dave2013, Thompson2014, Lagos2015,Hu2016} and observations \citep[e.g.,][]{Savage1977,Reach1994,Rachford2002,Gillmon2006,Lee2012,Balashev2014,Noterdaeme2015,Nakanishi2016}.
Analytic treatments of the \tra transition have been presented by \citet{Krumholz2008}, \citet{McKee2010} and 
 \citet{Sternberg2014} using a Str{\"o}mgren type analysis for the total steady state column density of \ho that is maintained by an incident photodissociating flux. 
In particular,  \citet[][hereafter, \citetalias{Sternberg2014}]{Sternberg2014} derived a scaling relationship for the total \ho column density in optically thick uniformly irradiated slabs as a function of the far-UV flux, the gas density, the dust absorption cross section and the \hh formation rate coefficient.  
% studied the 
%geometry (as opposed to a planar slab) for a cloud embedded in an isotropic radiation field.
\citet[][hereafter \citetalias{Bialy2016}]{Bialy2016} presented an analytic procedure for generating atomic ($\ho$) to molecular ($\hh$) density profiles for optically thick hydrogen gas clouds in Galactic star-forming regions.
These studies thus far have been instrumental in interpreting emission line
observations of \ho/\hh interfaces \citep{Lee2012, Bialy2015, Burkhart2015, Bihr2015,Bialy2017,Maier2017}, for estimating star-formation thresholds in external galaxies \citep{Leroy2008,Lada2012, Clark2014,Bialy2016,Burkhart2016},
 and for sub-grid components in hydrodynamics simulations \citep{Pelupessy2006, Thompson2014, Tomassetti2015}.

\begin{figure*}[t]
	\centering
	\includegraphics[width=1\textwidth]{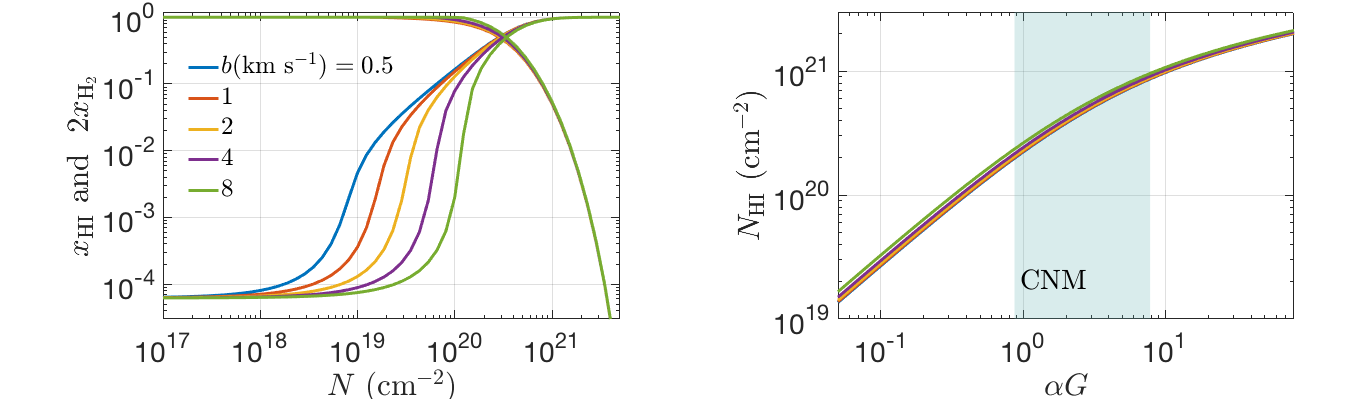} %/Turb//plots_for_paper/profiles_and_N1_tot_vs_b
	\caption{
Left: The H{\scriptsize I} and \hh  profiles, $x_{\ho} \equiv n_{\ho}/n$ and $2x_{\hh} \equiv 2n_{\hh}/n$, as functions of the column density $N$ (cloud depth), assuming $\phi_{g} Z'=1$ and $\alpha G=2$. Right: The total H{\scriptsize I} column density $N_{\ho}\equiv \int_0^{\infty}x_{\ho} dN$, as a function of $\alpha G$ as given by Eq.~(\ref{eq: N1tot}) assuming $\phi_{g} Z'=1$.
The CNM range for $\alpha G$ is  indicated.
Both panels are for a uniform-density optically thick slab irradiated by external beamed radiation field, and assuming various values of the Doppler line-broadening parameter $b_D$. Note that $N_{\ho}$ is insensitive to the choice of $b_D$.
		}
		\label{fig: N1_tot}
\end{figure*}
%But what about the density fluctuations we know are present in clouds?

% * <sb2580@gmail.com> 2016-07-24T15:43:45.039Z:
%
% We have to refer (here or elsewhere) to Glover Federrath and Klessen (2011)
%
% ^ <sb2580@gmail.com> 2016-07-24T15:47:21.104Z:
%
% http://adsabs.harvard.edu/cgi-bin/bib_query?arXiv:1103.3056 
%
% ^ <sb2580@gmail.com> 2016-07-24T15:48:02.149Z:
%
% and Glover & MacLow (2006) http://adsabs.harvard.edu/cgi-bin/bib_query?arXiv:astro-ph/0605121
%
% ^.

Despite the progress towards an analytic theory for the physics of the \tra transition, no current theory includes realistic turbulent density fluctuations.
% which are known to be present in both the molecular medium as well as in the neutral hydrogen.  
% This is potentially problematic because clouds on the scales of the $\ho/\hh$ transition are subject to supersonic turbulence, which creates strong density fluctuations.
The turbulent nature of molecular clouds is evident from a variety of observations including
non-thermal broadening \citep[][]{Stutzki1990, Dickey2001, Heiles2003, Heyer2004}, velocity/density power spectrum
\citep{Stanimirovic2001, Swift2008, Chepurnov2009, Pingel2013, Chepurnov2015},
%, Pingel2016
and fractal and hierarchical structures 
\citep[][]{Elmegreen1983, Vazquez-Semadeni1994, Burkhart2013}. 
Simulations and observations have shown that supersonic
turbulence creates filaments and regions of high density contrast
\citep{Kowal2007, Burkhart2009, Federrath2010}. 
 This behavior suggests that the assumption of uniform density in current analytic models for the \tra transition should be revisited.

The effects of turbulence on the chemical structure of interstellar clouds has been studied from various perspectives.
\citet{Xie1995}, \citet{Willacy2002} and \citet{Bell2011} studied the effects of turbulent mixing of chemical species through a diffusion approximation.
They found that atomic abundances (e.g., H, C, and O) may be significantly increased in the interiors of molecular clouds if the diffusion coefficient is large.
\citet{Levrier2012} studied the chemical structure of turbulent photodissociation regions (PDRs) using a post-processing approach, and found that the abundances of various molecules (e.g., H$_2$, CO, CH, and CN) exhibit strong deviations from a homogeneous PDR model \citep[cf.,][]{Offner2013}.
\citet{Glover2007b}, \citet{Glover2010}, \citet{Micic2011} and \citet{Valdivia2016} performed MHD simulations and followed the time-dependent H$_2$ formation self-consistently. 
They focused on the molecular content and showed that strong density compressions created by supersonic turbulent flows produce H$_2$ rapidly on timescales of few Myrs.

In this paper we study the effects of turbulent density perturbations on the $\ho/\hh$ structure in PDRs, focusing on the atomic gas produced by photodissociation at the cloud boundaries.
As we show, this gas is particularly useful for constraining the nature of turbulence, via 21 cm observations.
We consider a twofold approach, (i) via numerical MHD simulations supplemented by $\hh/\ho$ chemical balance calculation, and (ii) analytic modeling, and introducing a novel method for constraining the Mach number and the turbulence driving scale.

The paper is organized as follows: 
In \S \ref{sec: Theory} we provide a basic overview of \tra theory for uniform-density gas, as presented by \citetalias{Sternberg2014} and \citetalias{Bialy2016}.
In \S \ref{sec: den flucs} we discuss the effect of density fluctuations and the validity of our chemical steady-state  assumption when considering turbulence.
In \S \ref{sec: sims} we present the results of our MHD simulations.
% and discuss the arising volume density PDF, and the ``average along sightlines density PDF", which play an important role in the following sections.
In \S \ref{sec: H-H2 turb gas} we present computations of $\ho/\hh$ profiles and \ho column density PDFs for turbulent media.
In \S \ref{sec: model} we develop an analytic model for the \ho column density distribution.
In \S \ref{sec: observations} we demonstrate how our analytic model may be used to constrain turbulent parameters from 21 cm observations.
We discuss and summarize our results in \S \ref{sec: discussion}.

\section{Uniform density gas}
\label{sec: Theory}
In this section we review briefly the theory of \tra transition in steady-state, uniform density gas.
For a through discussion we refer the reader to \citetalias{Sternberg2014} and \citetalias{Bialy2016}.

%\subsection{The HI column for homogeneous gas}
%\label{sub: hom NH1}
%For a gas slab irradiated by external UV radiation field,
At any cloud depth, and for unidirectional radiation normal to the cloud surface, H$_2$ formation-destruction equilibrium is given by
\begin{equation}
\label{eq: H2 form_dest}
 R n \ n_{\ho} \ =  \ \frac{1}{2} D_0 \ f_{\rm shield}(N_{\rm H_2}) \mathrm{e}^{-\sg N} \ n_{\hh} \ ,
\end{equation}
%This Equation holds at any cloud depth 

where $R$ (cm$^3$~s$^{-1}$) is the H$_2$ formation rate coefficient, $n = n_{\ho}+2n_{\hh}$ is the total (atomic plus molecular) hydrogen volume density and $D_0$ (s$^{-1}$) is the free space H$_2$ photodissociation rate. In this expression, $f_{\rm shield}$ is the H$_2$ self-shielding function that depends on the H$_2$ column density $N_{\hh}$ and also on the absorption-line Doppler broadening parameter $b_{D}$ (km~s$^{-1}$) (\citealt{Draine1996}, \citetalias{Sternberg2014}). The factor $\mathrm{e}^{-\sg N}$ is the dust absorption attenuation term, where $\sg$ (cm$^{2}$) is the dust-grain absorption cross section per hydrogen nuclei integrated over the Lyman-Werner dissociation band (11.2-13.6 eV; hereafter LW-band), and $N=N_{\ho}+2N_{\hh}$  is the total, atomic plus molecular column density. 
The factor of 1/2 accounts for absorption of half the radiation by the optically thick slab.

Assuming that all of the photodissociating radiation is absorbed in the cloud, the total \ho column density, converges to a finite value, $N_{\ho}$.
As shown by \citetalias{Sternberg2014}, assuming  slab geometry, and for constant density
\begin{align}
\label{eq: N1tot}
N_{\ho} \ &= \ \frac{1}{\sg} \ \ln \left[ \frac{\alpha G}{2} \ + \ 1 \right] \\ \nonumber
&= 5.3 \times 10^{20} \frac{1}{\phi_{g} Z'} \ \ln \left[ \frac{\alpha G}{2} \ + \ 1 \right] \ {\rm cm^{-2}}
\end{align}
where in the second equality 
\begin{equation}
\sg = 1.9 \times 10^{-21} \phi_{g} Z' \ {\rm cm^2}
\end{equation}
where $Z'$ is the dust-to-gas ratio relative to Galactic, and $\phi_{g}$ is a factor of order unity that characterizes the dust absorption properties.  In Eq.~(\ref{eq: N1tot}), $\alpha \equiv D_0/(Rn)$ is the (dimensionless) ratio of free-space H$_2$ photodissociation and H$_2$ formation rates. The dimensionless factor $G \equiv \sg  \int f_{\rm shield}(N_{\hh}) \mathrm{e}^{-2\sg N_{\hh}} dN_{\hh}$ is ``the effective shielding factor" and may be expressed as $G = 3.0 \times 10^{-5} \phi_{g} Z' (9.9/[1+8.9\phi_{g} Z'])^{0.37}$ (\citetalias{Bialy2016}). 
 %, and the superscript ``hom" stands for ``homogeneous".
The combination
\begin{equation}
\label{eq: aG}
\aG = \frac{D_0 G}{R n} = 2.0 \ I_{\rm UV} \left( \frac{30 \ {\rm cm^{-3}}}{n} \right)  \ ,
\end{equation}
has the physical meaning of an effective dissociation parameter taking into account H$_2$-shielding and the competition with dust absorption. The numerical value in Eq.~(\ref{eq: aG}) is for $R = 3 \times 10^{-17}$~cm$^3$~s$^{-1}$, $\phi_{g} Z'=1$, and $D_0=5.8 \times 10^{-11}\ I_{\rm UV}$~s$^{-1}$ (\citetalias{Sternberg2014}) where $I_{\rm UV}$ is the radiation strength relative to the \citet{Draine1978} field.
The product \aG may be small or large for realistic astronomical environments.
For example, for the star-forming region W43 and for the Perseus molecular cloud, \citet{Bialy2017} and \citet{Bialy2015} derived $\aG  \sim 20$ and $\aG \sim 10$, respectively, whereas for a sample of dwarf irregular galaxies in the LITTLE THINGS survey \citep{Hunter2012}, \citet{Maier2017} deduced $\aG < 1$.
For cold neutral medium (CNM) which is in pressure equilibrium with the warm neutral medium (WNM), the $n/I_{\rm UV}$ ratio is restricted to the range $\approx 8-70$~cm$^{-3}$ \citep{Wolfire2003}, giving $(\aG)_{\rm CNM} \approx 1-8$.

The left panel of Fig.~\ref{fig: N1_tot} shows the  \ho and $\hh$ fractional abundance profiles, $x_{\ho} \equiv n_{\ho}/n$ and $2 x_{\hh} \equiv 2n_{\hh}/n$, as functions of cloud depth, as parameterized by the gas column density $N$. The various curves are for $\phi_{g} Z'=1$, $\aG=2$ and Doppler parameter ranging from $b_{D}=0.5$ to 8 km s$^{-1}$. 
With increasing depth the radiation is absorbed by H$_2$ photodissociation events and by dust absorption, and the gas makes the transition from atomic to molecular form. 
For larger $b_{D}$, the Doppler cores of the H$_2$ lines are broader, and the onset of self shielding occurs at larger cloud depths. 
However, the \tra transition point is insensitive to $b_D$ because it occurs deeper in the cloud where the LW-flux is absorbed in the H$_2$ damping wings (\citetalias{Bialy2016}).
% Detailed \tra density profiles for various values of $\aG$ and $\sg$ are presented in \citetalias{Bialy2016} (see their Figs.~4-6).

The right panel of Fig.~\ref{fig: N1_tot} shows the total \ho column density, $N_{\ho}$, as a function of $\aG$.
For large $\aG$ (``the strong field limit") dust absorption determines the \ho column density, and $N_{\ho}$ is weakly dependent on $\aG$.
% The strong dust absorption quickly removes any trace UV photons after the \tra transition, and the \ho column saturates to the asymptotic value $N_{\ho}$ shortly after the \tra transition. 
For small \aG (``the weak field limit"), 
H$_2$ self-shielding dominates and $N_{\ho} \propto \aG$.
As for the \tra transition points, $N_{\ho}$ is insensitive to $b_{D}$ (and for the same reason). 
% Again, this is because $N_{\ho}$ accumulates deeper in the cloud, where absorption is dominated by the damping wings. 
The insensitivity of $N_{\ho}$ to the Doppler parameter is important in our analysis of turbulent media and the line broadening induced by turbulent motions.

% Because the \ho column saturates deep in the cloud where the radiation is absorbed by the H$_2$-line damping wings, $N_{\ho}$ is insensitive to the value of the Doppler parameter.
% The gradual absorption by H$_2$ lines allows trace UV photons to penetrate into the predominantly H$_2$ regions, deep after the \tra transition.
% As a result, the \ho column continues to accumulate within the H$_2$ region, all the way until dust absorption kicks in (at $N \sim 1/\sg \approx 5\times 10^{20}$~cm$^{2}$).

\section{Density Fluctuations and Timescales}
\label{sec: den flucs}
As discussed above, two basic assumptions for the \tra density profiles and the \ho column density (as shown in Fig.~\ref{fig: N1_tot} and given by Eq.~\ref{eq: N1tot}), are a constant gas density $n$ and chemical steady state. 
In this paper we relax the assumption of constant density by considering turbulent density fluctuations, while retaining the assumption of chemical steady state.
Density fluctuations are naturally produced in the cold ISM by supersonic turbulence.
Since the H$_2$ formation-to-removal rate ratio is proportional to density, a local increase in density increases the local H$_2$ fraction. 
Such a perturbation also affects deeper locations in the cloud through enhanced H$_2$ self-shielding which depends on the H$_2$ column density.

For a given density $n$, and a local (attenuated) dissociation rate $D$, the chemical time is 
\begin{equation}
t_{\rm chem}=\frac{1}{2Rn+D} \ \ \ .
\end{equation}
% \begin{equation}
% t_{\rm chem} \ = \ \frac{1}{2Rn+D} \ ,
% \end{equation}
In the outer atomic layers, $D \gg 2Rn$ and $t_{\rm chem} \simeq 1/D$ is the photodissociation time, which is typically very short.
For example, for unshielded gas, $D=D_0/2=2.9 \times 10^{-11}$~s$^{-1}$ and $t_{\rm chem}=1.1\times10^3$~yr.
Beyond the atomic-to-molecular transition, $D<2Rn$ and $t_{\rm chem} \simeq 1/(2Rn)$ is the H$_2$ formation time, which can become long.
The strongest effect of density perturbations will occur near the \tra transition points where $Rn$ and $D$ are comparable.
At the transition point,  $D=2Rn$, and 
\begin{equation}
\label{eq: t_chem}
t_{\rm chem} = \ \frac{1}{4Rn} \ \approx 2.7 \frac{1}{\phi_R Z' T_2^{1/2} n_2} \ {\rm Myr} \ ,
\end{equation} where the rate coefficient
\begin{equation}
R= 3 \times 10^{-17} \ T_2^{1/2} \ \phi_R \ Z' \ {\rm cm^3 \ s^{-1}} \ ,
\end{equation}
Here, $T_2 \equiv T/10^2$~K, $n_2 \equiv n/10^2$~cm$^{-3}$,  and $\phi_R$ is a  factor of order unity.
Typically, $\phi_R \approx 1$, however, in some environments the H$_2$ formation rate may be enhanced.
For example, \citet{Habart2003, Habart2004} found that moderately illuminated PDRs, such as Oph W, S140 and IC 63,  may have $\phi_R \approx 5$, considerably reducing the chemical time.

Irrespective of any density fluctuations,
the chemical time must be short compared to the cloud lifetime, $t_{\rm cloud}$.
This requires 
\begin{equation}
n \geq \frac{1}{4Rt_{\rm cloud}} \approx 27 \frac{1}{\phi_R Z' T_2^{1/2}} \left(\frac{10 \ {\rm Myr}}{t_{\rm cloud}} \right) \ {\rm cm^{-3}} \ ,
\end{equation}
where we have normalized $t_{\rm cloud}$ to characteristic lifetime of 10 Myrs. 

In a turbulent medium, we also require $t_{\rm chem} \leq t_{\rm turb}$, where $t_{\rm turb}$ is the characteristic time over which turbulent density fluctuations are formed and destroyed.
The turbulent time is
\begin{equation}
\label{eq: t_turb basic}
t_{\rm turb} \simeq \frac{L_{\ho}}{\delta \varv_{1d}(L_{\ho})}
\end{equation}
where
\begin{equation}
\label{eq: HI length-scale}
L_{\ho} \ = \ \frac{1}{\sigma_g n} = 1.7 \ \frac{1}{\phi_{g} Z'  n_2} \ {\rm pc} \ ,
\end{equation} is the characteristic length-scale of the \ho layer, and $\delta \varv_{1d}(L_{\ho})$ is the 1d velocity dispersion over  $L_{\ho}$.
Following the linewidth-size relation \citep[][]{Larson1981,McKee2007}, the 3d velocity dispersion over a length-scale $\ell$ is
\begin{equation}
\label{eq: linewidth size relation}
\delta \varv(\ell) = \delta \varv(L_{\rm drive}) \left( \frac{\ell}{L_{\rm drive}} \right)^{1/2} \  \ \ (L_s \leq \ell \leq L_{\rm drive}) \ ,
\end{equation}
where $L_{\rm drive}$ is the outer driving scale, and $L_s$ is the sonic length for which  $\delta \varv(L_s) \equiv c_s$, where $c_s$ is the sound speed. 
Defining the Mach number $\ms \equiv \delta \varv(L_{\rm drive})/c_s$, and assuming an isotropic velocity field, we get
\begin{align}
\label{eq: t_turb}
t_{\rm turb} &= \frac{\sqrt{3}}{c_s \ms} \ \frac{1}{\sg n} \left( \frac{L_{\ho}}{L_{\rm drive}} \right)^{-1/2} \\ \nonumber
&\approx 4.0 \frac{1}{\ms T_2^{1/2} n_2 \phi_g Z'} \left( \frac{L_{\ho}}{L_{\rm drive}} \right)^{-1/2} \ {\rm Myr} \ ,
\end{align} 
where in the second equality we used  $c_s=0.72T_2^{1/2}$~km~s$^{-1}$, assuming a mean particle mass of 1.6 the proton mass, at the transition point where $x_{\ho}=2x_{\hh}$, and including Helium.

\begin{figure*}
	\centering	\includegraphics[width=1\textwidth]{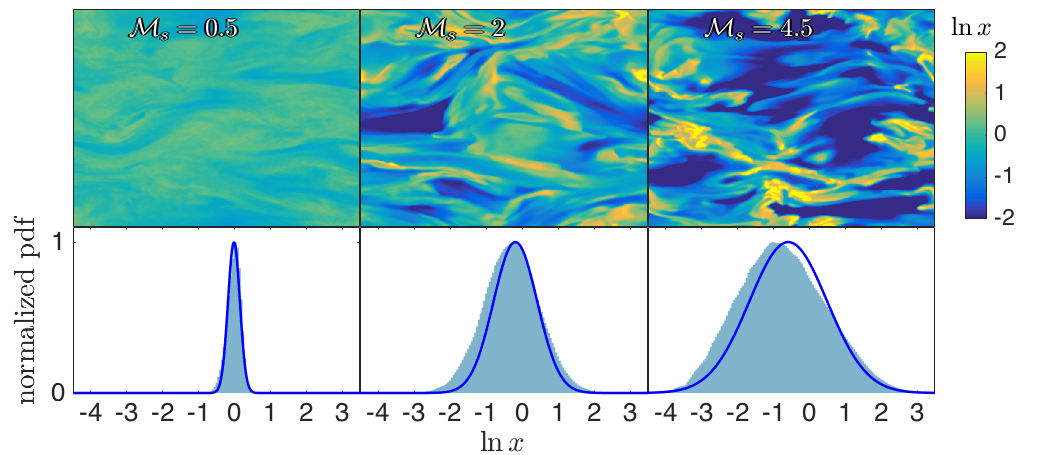}
	\caption{
	Top: slices of $\ln x \equiv \ln \ n/\langle n \rangle$ from the $\ms=0.5, 2$ and $4.5$ simulations. Bottom: The probability distribution functions of $\ln x$ for the entire simulation box.
The blue curve is the normal distribution assuming the  $\sigma_{\ln x} - \ms$ relation (Eq.~\ref{eq: s-M relation lnx}) with $b=1/3$.}
\label{fig: maps}
%generated with Turb/plots_for_paper/plot_slices_and_histograms
\end{figure*}

The ratio of the chemical and turbulent times is then
\begin{equation}
\label{eq: t_chem t_turb ratio}
\frac{t_{\rm chem}}{t_{\rm turb}} \approx 0.66 \ms \frac{\phi_g}{\phi_R} \left(\frac{L_{\ho}}{L_{\rm drive}} \right)^{1/2} \ .
\end{equation}
Eqs.~(\ref{eq: t_turb}) and (\ref{eq: t_chem t_turb ratio}) are for $L_s \leq L_{\ho} \leq L_{\rm drive}$.
For $L_{\ho}>L_{\rm drive}$, $\delta \varv \rightarrow \delta \varv(L_{\rm drive})$, and $L_{\ho}/L_{\rm drive}$ should be replaced with unity. For $L_{\ho}<L_s$ density perturbations are negligible as they are smoothed out by pressure waves.
At  $L_{\ho} = L_s$, $L_{\ho}/L_{\rm drive}=1/\ms^2$. 
Thus, for moderate Mach numbers, $t_{\rm chem}/t_{\rm turb}$ remains close to unity, possibly ranging from $0.66 (\phi_g/\phi_R)$ to $0.66 (\phi_g/\phi_R) \ms$.
Throughout this paper we assume chemical equilibrium. We will consider the more complicated time-dependent problem elsewhere.

Turbulent motions may also affect the $\ho$/$\hh$ structure by shifting the frequencies of the \hh absorption lines, and reducing the efficiency of H$_2$-self shielding.
This affects the $\ho$/$\hh$ profiles at intermediate depths, $10^{14} \leq N_{\hh} \leq 10^{18}$~cm$^{-2}$ where absorption is dominated by the Doppler cores \citep[][]{Gnedin2014}.
However, since most of the \ho gas is accumulated at greater cloud depths where the radiation is absorbed in the H$_2$-line damping wings, the velocity shifts do not affect $N_{\ho}$ (see the discussion in \S \ref{sec: Theory}).
The Doppler broadening is important for optically thin medium  (to the LW-radiation), but not to optically thick clouds that have fully converted $\tra$.
Since our focus is on optically thick gas, we assume a constant $b_D=2$~km~s$^{-1}$ throughout our calculations, and do not include any corrections to the Doppler parameter or the H$_2$ self-shielding function  \citep[e.g.,][]{Gnedin2014}.

\section{MHD Simulations}
\label{sec: sims}

In this Section, we use MHD simulations to obtain realistic density profiles for sub- and supersonic turbulent gas. 
Our simulations are isothermal and non-self gravitating \citep[cf.][]{Jappsen2005}.
This allows a natural extension of the \citetalias{Sternberg2014} \tra transition model, which is inherently isothermal, into the turbulent regime.
Furthermore, an isothermal equation of state (EOS) allows a simple estimate of the density dispersion (Eq.~\ref{eq: s-M relation}, below).
In reality, the \tra transition takes place in a non-isothermal medium with heating and cooling processes acting, e.g., depth dependent photoelectric heating versus [C{\small II}] emission line cooling \citep{Tielens1985, Sternberg1989}. 
However, for moderate sonic Mach numbers ($\ms \lesssim 5$),  the density and column density PDFs, are similar in simulations of isothermal or non-isothermal EOS \citep[e.g.,][]{Glover2007b,Federrath2015a}.
% Therefore, it is the supersonic nature of the medium, rather than the EoS or thermal instability, that will dominate the PDF at such Mach numbers. 

% In non-dimensional form, these are
% \begin{align}
%  \frac{\partial x}{\partial \bar{t}} + \bar{\nabla} \cdot (x {\bf u}) &= 0, \\
%  \frac{\partial (x {\bf u})}{\partial \bar{t}} + \bar{\nabla} \cdot \left[ \rho {\bf u} {\bf u} + \left( \ms + 2\pazocal{M}_A^2  \right) {\bf I} - \pazocal{M}_A^2 {\bf b}{\bf b} \right] &= \bar{\bf f}, \\ 
%  \frac{\partial {\bf b}}{\partial \bar{t}} - \bar{\nabla} \times ({\bf u} \times{\bf b}) &= 0 \ ,  
% \end{align}
We use a third-order-accurate hybrid essentially nonoscillatory scheme \citep{Cho2002} to solve the ideal MHD equations,
\begin{align}
 \frac{\partial \rho}{\partial t} + \nabla \cdot (\rho \pmb{\varv}) &= 0, \\
 \frac{\partial \rho \pmb{\varv}}{\partial t} + \nabla \cdot \left[ \rho \pmb{\varv} \pmb{\varv} + \left( p + \frac{B^2}{8 \pi} \right) {\bf I} - \frac{1}{4 \pi}{\bf B}{\bf B} \right] &= {\bf f},  \\
 \frac{\partial {\bf B}}{\partial t} - \nabla \times (\pmb{\varv} \times{\bf B}) &= 0 \ ,
\end{align}
where $\rho$ is density,
% \footnote{In the remainder of the text we denote the normalized volume density as $x\equiv n/\langle n \rangle$, where $\langle n \rangle$ is the mean volume density.}
${\bf B}$ is magnetic field, $p$ is the gas pressure, ${\bf I}$ is the identity matrix and $\bf{f}$ is the specific force.
We assume zero-divergence condition $\nabla \cdot {\bf B} = 0$,  periodic boundary conditions, and an isothermal equation of state $p = c_s^2 \rho$.
For the source term $\bf{f}$, we assume a random large-scale solenoidal driving at a wave number $k\approx 2.5$ (i.e.~1/2.5 the box size). 
The simulations have 512$^3$ resolution elements and have been employed in many previous works
\citep{Cho2003, Burkhart2009, Burkhart2010,
Kowal2007, Kowal2009, Kowal2011}.

Each simulation is defined by the sonic Mach number $\ms \equiv  |\pmb{\varv}|/c_s$, and the Alfv\'{e}nic Mach number $\pazocal{M}_A \equiv |\pmb{\varv}|/ \langle \varv_A \rangle$, where $\pmb{\varv}$ is the velocity, $c_s$  and $\varv_A$ are the isothermal sound speed and the Alfv\'en speed, and $\langle \cdot \rangle$ denotes averages over the entire simulation box.
We show results for $\ms=0.5$, 2, and 4.5 simulations, i.e., subsonic, transonic, and supersonic gas.
As we show below, the value of the sonic Mach number strongly affects the variance of the density field. 
The simulations are sub-alfv\'{e}nic with $\pazocal{M}_A=0.7$ (i.e.~strong magnetic field).
We have also considered super-Alfv\'{e}nic ($\pazocal{M}_A=2.0$) simulations and found that the results are weakly sensitive to the value of $\pazocal{M}_A$. Because the simulations are non self-gravitating they are scale-free and we may assign any desired physical scale for the box length and density (see \citealt[][Apendix]{Hill2008}).
 In this section we keep the results general and do not apply any physical scaling to the simulations. 
We scale the simulations to physical units in \S \ref{sec: H-H2 turb gas} below.

% The simulations are non self-gravitating and are scale-free.

% In non-dimensional form, the MHD equations take the form
% Each resolution element is then $=L/512 \approx 0.06$~pc, well within the sonic scale-length.
% Thus, the smallest-scale density perturbations are resolved in our simulations.
%, with $x$ being often $\ll$ or $\gg 1$.
%Next we present and discuss our numerical results and compare them to the analytic model in \S \ref{sec}.

% \subsection{Scaling the Simulations to Physical Units}

% We emphasize that the simulations used in this work are performed in dimensionless units.  When appropriate, these simulations can be rescaled to physical units.
% For example, in order to scale the box to a physical length 
% we can define a length scaling factor
% \begin{equation}
% x_0 = \frac{L_\mathrm{box}}{N_x}.
% \end{equation}
% where $L_\mathrm{box}$ is the physical size of the cloud (e.g. in parsecs) and  the  total volume of the simulation in dimensionless units is $N_x \times N_y \times N_z$, where for the simulations considered here $N_x=N_y=N_z$. 

% Similarly the velocity field in the simulation can be scaled  from relating the sound speed in code units to physical units, 

% \begin{equation}
% v_0 = \frac{c_{s,\obs}}{\tilde{c_{s}}}
% \end{equation}
% where the physical sound speed of an ideal gas is
% \[
% c_{s,\obs} = \sqrt{\frac{\partial p}{\partial \rho}} = \sqrt{\frac{k_B T_\obs}{\mu}},
% \]
% where $T_\obs$ is the temperature and $\mu$ is the mean particle mass. For a full discussion regarding the scaling of these MHD simulations see the Appendix of \citet{Hill2008}.  

\subsection{The 3D Density Distribution}
Fig.~\ref{fig: maps} shows three random density cuts (upper panels) through the $\ms=0.5, 2$ and 4.5 simulations. 
The color axis corresponds to $\ln x$ where $x \equiv n/\langle n \rangle$.
The density is nearly uniform for the subsonic simulation, but once the Mach number exceeds unity strong density fluctuations are generated.
The lower panels show the PDFs of $\ln x$ for these simulations (shaded) .
The $\ln x$ distributions are nearly Gaussian with a standard deviation that increases with Mach number. 
% the probability distribution function of density perturbations induced by turbulence, is well described by the lognormal distribution,
% \begin{equation}
% \label{eq: lognormal PDF}
% \frac{df}{dx} \ = \ \frac{1}{\sqrt{2 \pi} \sigma_{\ln x} \ x} \ \exp \left[ -\frac{(\ln x - \mu_{\ln x})^2}{2\sigma_{\ln x}^2} \right] \ .
% \end{equation}
% Here $x\equiv n/\langle n \rangle$ is the normalized volume density and $\langle n \rangle$ is the mean volume density.
% $\sigma_{\ln x}$ and $\mu_{\ln x}$ are the standard deviation and mean of $\ln x$, related through $\mu_{\ln x}=-0.5\sigma_{\ln x}^2$.
This is in agreement with previous 
studies \citep[e.g.,][]{Padoan1997, Passot1998, Federrath2008, Price2011,Burkhart2012,Molina2012} that found that $x$ is lognormally distributed (and $\ln x$ is Gaussian), with
\begin{align}
\label{eq: s-M relation}
\sigma_{x} & \simeq  b \ms \\
\sigma_{\ln x}^2 &  \simeq  \ln[1+(b \ms)^2] \ ,
\label{eq: s-M relation lnx}
\end{align} In these expressions $\sigma_x$ and $\sigma_{\ln x}$ are the standard deviations of the $x$ and $\ln x$ distributions, and where $x$ has a unit mean (by definition) and the mean of $\ln x$ is $\mu_{\ln x} = -(1/2) \sigma_{\ln x}^2$. 
The proportionality constant $b$ depends on the nature of the turbulent driving, and ranges from $1/3$ to $1$ for for pure solenoidal or compressive driving respectively \citep{Nordlund1999, Federrath2008, Federrath2010}.
The solid curves in Fig.~\ref{fig: maps} are Gaussians with $\sigma_{\ln x}$ as given by Eq.~(\ref{eq: s-M relation lnx}) with $b=1/3$, appropriate for our solenoidly driven simulations.
The agreement is not perfect due to small deviations from the phenomenological $\sigma - \ms$ relation, given by Eq.~(\ref{eq: s-M relation lnx}).

% This is further shown in Fig.~\ref{fig: sx-Ms relation} where we plot $\sigma_x$ as a function of $\ms$ for four simulations of $\ms=0.5, 2, 4.5$, and 7, along with the $\sigma_x=b\mathcal{M}_s$ relation ($b=1/3)$.

\begin{figure}
	\centering	\includegraphics[width=0.5\textwidth]{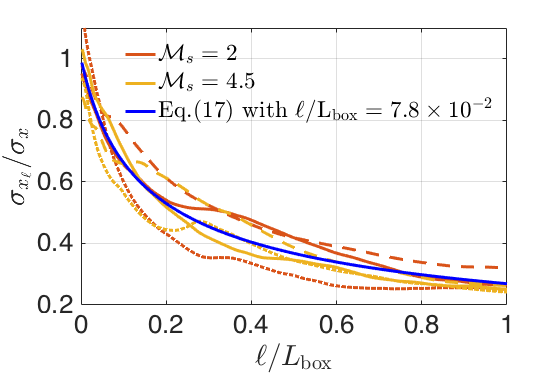}
    %generated with /Turb/plots_for_paper/sig_L
	\caption{
The ratio of the standard deviations of $x_{\ell}$ and $x$ (see text), as a function of the averaging length ${\ell}/L_{\rm box}$.
Results for the $\ms=2.0$, and 4.5 simulations, for LOS along the $X$ (solid), $Y$ (dashed) and $Z$ (dotted) directions  are shown.
The blue curve is the theoretical relation (Eq.~\ref{eq:sigxl_2}) with the best fitted parameter $L_{\rm dec}/ L_{\rm box}= 7.8 \times 10^{-2}$.	
}
\label{fig: sig_L}
\end{figure}

\subsection{Line-of-sight-averaged Densities}
\label{sub: LOS averages sims}
We now discuss an important distribution that will play a crucial role in determining the \ho column density PDF. 
For a column of length $\ell$ we define the average density along a line-of-sight (LOS)
\begin{equation}
\label{eq: xLdefinition}
x_{\ell} \equiv \frac{\int_0^{\ell} x \ \mathrm{d}{\ell}'}{\ell} \ \ \ ,
\end{equation} where $0 \leq \ell \leq L_{\rm box}$, and where $L_{\rm box}$ is the simulation box length. 
For any given $\ell$, different sightlines have different density profiles, and thus the set $x_{\ell}$ form a random variable. We refer to the distribution of $x_{\ell}$ as the ``LOS averaged density distribution".

For turbulent cascade the density is correlated over all scales up to the driving scale. However, the correlation decreases with increasing spatial separation \citep[][hereafter \citetalias{VazquezSemadeni2001}]{VazquezSemadeni2001}. 
To obtain an analytic description for the $x_{\ell}$ distribution, we assume that the correlation may be described with a single parameter, $L_{\rm dec}$, hereafter the ``decorrelation scale'', such that for ${\ell} < L_{\rm dec}$ the density is effectively constant (i.e., maximally correlated) and for $\ell \geq L_{\rm dec}$ the density cells are uncorrelated. 
The number of independent density cells along a LOS of length $\ell$ is then
\begin{equation}
\label{eq: paz N}
\pazocal{N}(\ell) = 1+\frac{\ell}{L_{\rm dec}} \ \ \ ,
\end{equation}
and the $x_{\ell}$ distribution may be viewed as the sampling distribution of the mean, for which
\begin{equation}
\label{eq: sigma xL}
\sigma_{x_{\ell}} \simeq \frac{\sigma_x}{\sqrt{\pazocal{N}(\ell)}} \  \ \ .
% \simeq \frac{b \ms}{\sqrt{\pazocal{N}(\Delta z)}}  ,
\end{equation} 
This distribution is often encountered in the calculation of errors in repeated measurements \citep{Barlow1989}. 
For $\ell \ll L_{\rm dec}$, the LOS contains a single fluctuation $\pazocal{N} \approx 1$, and $\sigma_{x_{\ell}}=\sigma_x$. For ${\ell} \gg L_{\rm dec}$, $\pazocal{N} \gg 1$, the LOS contains many turbulent fluctuations, and $\sigma_{x_{\ell}} \ll \sigma_{x}$ as the fluctuations are averaged out.

\begin{figure}
	\centering	
    	\includegraphics[width=0.5\textwidth]{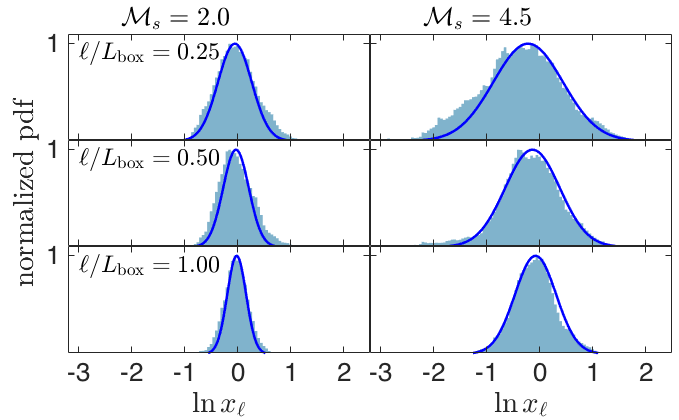}
    %generated with /Turb/plots_for_paper/sig_L
	\caption{	
Probability distribution functions of $\ln x_{\ell}$ for the $\ms=2.0$ and 4.5 simulations, and for ${\ell}/L_{\rm box}=0.2$, 0.5 and 1. The blue curves are Gaussians with standard deviations given by 
Eq.~(\ref{eq: s-M relation},\ref{eq: paz N}-\ref{eq: Ldec}).
}
\label{fig: xL_dists}
\end{figure}

Fig.~\ref{fig: sig_L} shows the ratio of the standard deviations, $\sigma_{x_{\ell}}/\sigma_x$, as a function of ${\ell}/L_{\rm box}$, as calculated for our $\ms=2$ and 4.5 simulations. We consider sightlines along the $X$ (solid), $Y$ (dashed), and $Z$ (dotted) directions.
The solid blue curve is a fit for the predicted relation
\begin{equation}
\label{eq:sigxl_2}
\frac{\sigma_{x_l}}{\sigma_x}=\left(1+\frac{\ell}{L_{\rm dec}}\right)^{-1/2} = \left(1+\frac{L_{\rm box}}{L_{\rm dec}}\frac{\ell}{L_{\rm box}}\right)^{-1/2} \ ,
\end{equation}
 with the best fitted parameter
\begin{equation}
\label{eq: Ldec}
\frac{L_{\rm dec}}{L_{\rm box}} = 7.8\times10^{-2} \ ,
\end{equation} (equivalent to 40 out of 512 cells).
Evidently, our simplified treatment for the density correlation gives a reasonable estimate for the $x_l$ dispersion.

\begin{figure*}
\centering
\includegraphics[width=1\textwidth]{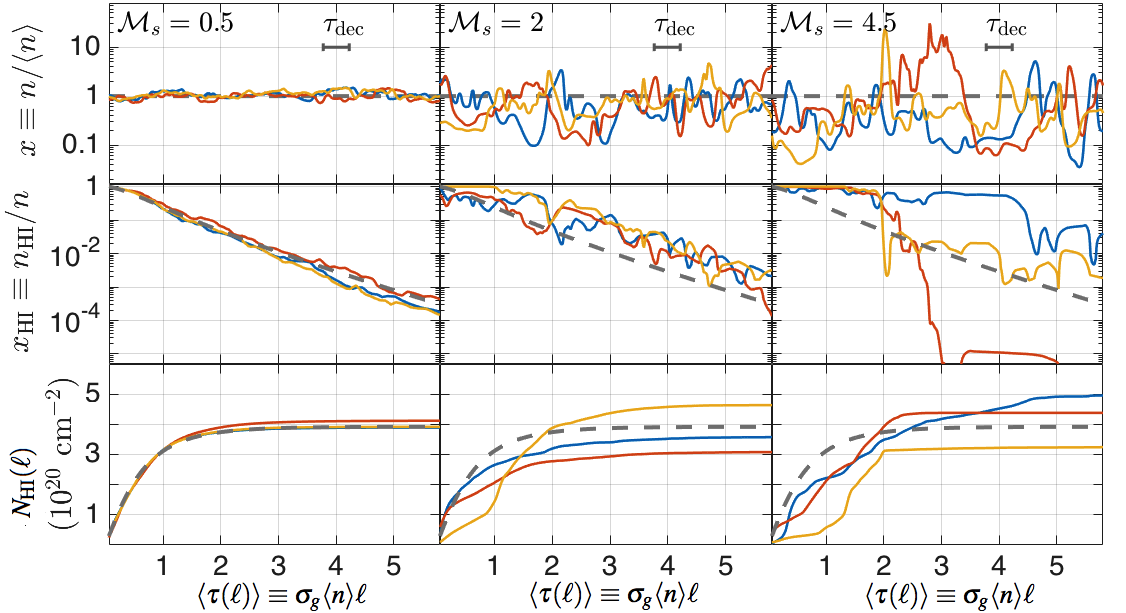}
%generated with /Turb/plots_for_paper/plot_profiles
	\caption{The normalized density profiles (top), the normalized H{\scriptsize I} profiles (middle), and the accumulated H{\scriptsize I} column density (bottom), as calculated for three arbitrary LOS through the $\ms=0.5$, 2 and 4.5 simulations, assuming $\overline{\aG}=2$. The cloud depth in the abscissa is in units of the mean H{\scriptsize I} scale, $1/(\sg \langle n \rangle)$. 
	The decorrelation width, $\tau_{\rm dec}=0.45$ is indicated by the horizontal bar, representing a typical length-scale for the density fluctuations. The homogeneous solutions are shown for comparison (dashed curves). }
	\label{fig: N1_tot_profiles}
\end{figure*}

In Fig.~\ref{fig: xL_dists} we show PDFs of $\ln x_{\ell}$  (shaded), for $\ms=2$ and 4.5 and ${\ell}/L_{\rm box}=0.25$, 0.5 and 1. 
The PDFs have distorted Gaussian shapes, becoming narrower with increasing ${\ell}$, as expected from Eq.~(\ref{eq:sigxl_2}).
% Assuming $\ln x_L$ is normally distributed, Eqs.~(\ref{eq: sigma xL}, \ref{eq: s-M relation}) predict 
% \begin{equation}
% \label{eq: sig_lnxL}
% \sigma_{\ln xL}^2=\ln \left[ 1+\frac{(b\ms)^2}{\pazocal{N}} \right] \ ,
% \end{equation}
The blue curves are Gaussians with standard deviations according to Eqs.~(\ref{eq: s-M relation}) and (\ref{eq: paz N}-\ref{eq: Ldec}).
We conclude that $x_{\ell}$ is indeed well described by a lognormal with $\sigma_{x_{\ell}}=b\ms/\sqrt{\pazocal{N}}$.
We note that this relationship derived for $\sigma_{x_{\ell}}$ is complimentary to the column density variance - \ms relationship derived in \citet{Burkhart2012} (their Equation 4 with $A=0.11$) in the limit that ${\ell}=L_{\rm box}$.
However the relationship present here is more general and provides a method to determine the driving scale and decorrelation scale via measurement of column density variance. 

It is instructive to write $L_{\rm dec}$ in terms of the driving scale.
For all our simulations, $L_{\rm drive}=L_{\rm box}/2.5$, and with Eq.~(\ref{eq: Ldec}) we get
\begin{equation}
\label{eq: L_dec L_drive}
\frac{L_{\rm dec}}{L_{\rm drive}} = 0.20 \ .
\end{equation}
The decorrelation scale is smaller than, but of order of the driving scale.
This is because the driving process introduces density (and velocity) correlations, which cascade down to smaller scales.
Eq.~(\ref{eq: L_dec L_drive}) is a general relation for $L_{\rm dec}$ and $L_{\rm drive}$, although the prefactor may depend on the driving details (e.g.~compressional versus solenoidal).
% Furthermore, since for typical CNM conditions the inertial range is large (a Reynolds number $Re \gg 1$; \citealt{McKee2007}), $l_{\rm dec}$ lies well within the inertial range; $l_{\nu} \ll l_{\rm dec} < l_{\rm drive}$, where $l_{\nu}$ is the dissipation scale, and $l_{\rm dec}$ is indeed expected to scale with $l_{\rm drive}$.
\citetalias{VazquezSemadeni2001} and \citet{Fischera2004} also studied the $L_{\rm dec}-L_{\rm drive}$ relation (using alternative methods) and obtained $L_{\rm dec} =0.33 L_{\rm drive}$ and $L_{\rm dec} =0.13 L_{\rm drive}$, respectively.
Our value for the $L_{\rm dec}-L_{\rm drive}$ relation is also in good agreement with that of \citet{Kowal2007}.

While $L_{\rm dec}<L_{\rm drive}$, it is typically larger than the sonic scale.
For example, Eqs.~(\ref{eq: linewidth size relation}) and (\ref{eq: L_dec L_drive}) suggest that $L_{\rm dec} \geq L_s$ as long as $\ms \geq 2.2$. This is important for our model, because $L_{\rm dec}$ represents the scale below which the density becomes effectively uniform. But if $L_{\rm dec}<L_s$, then $L_{\rm dec}$ should be everywhere replaced with $L_s$.

 \begin{figure*}
    \includegraphics[width=1\textwidth]{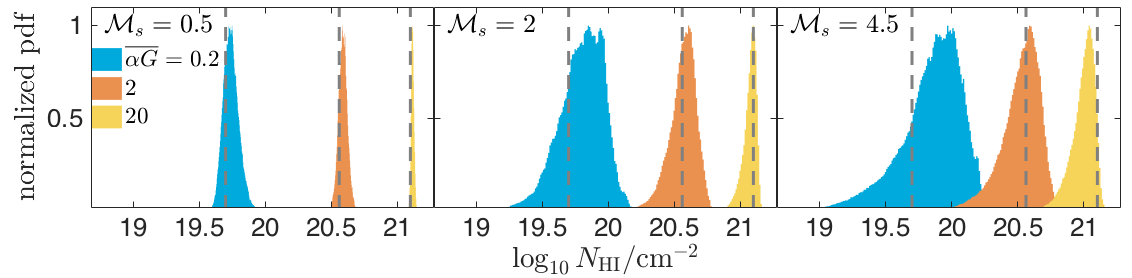}
        \includegraphics[width=1\textwidth]{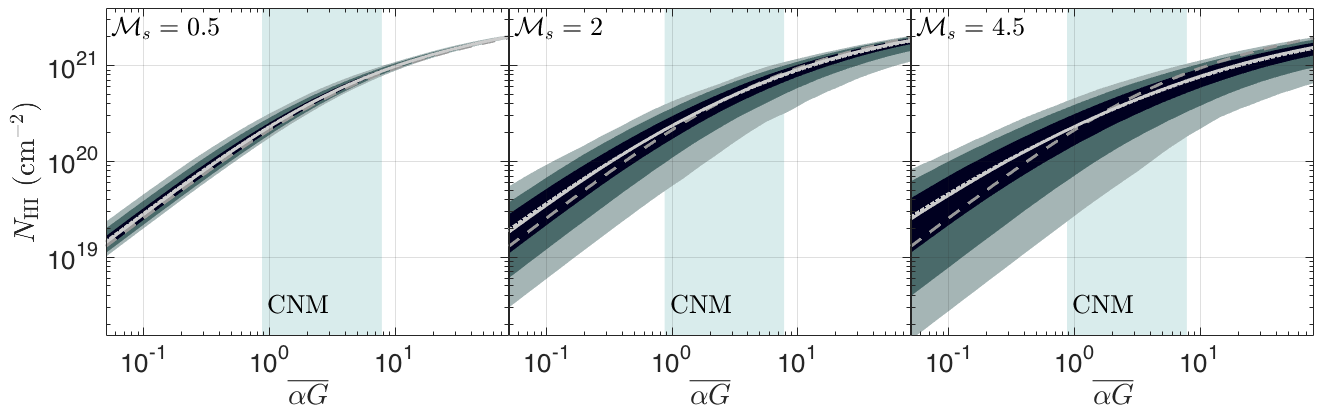}
    % Turb/plots_for_paper/plot_HI_PDFs, and N1_vs_aG
	\caption{Top: the PDFs of $\log_{10} N_{\ho}$, as calculated for the $\ms=0.5$, 2 and 4.5 simulations, assuming $\overline{\aG}=0.2, 2$ and 20. Bottom: the median (solid curves), mean (dotted - almost converges with the median), and the 68.3, 95.5, 99.7 percentiles (shaded regions) as functions of $\overline{\aG}$. 
	For both panels, the simulations assumed scale corresponds to  $\tau_{\rm dec}=0.45$.
 The dashed curves are the homogeneous solutions, Eq.~(\ref{eq: N1tot}), for comparison.
}	
\label{fig: N1_dists_sims}
\end{figure*}

\section{$\ho-to-\hh$ in Turbulent Gas}
\label{sec: H-H2 turb gas}
In this section we present atomic and molecular density profiles and integrated \ho column density distributions for non-homogeneous, turbulent gas.
We use the density field obtained from our MHD simulations, and assume a unidirectional UV flux incident of the box from one side.
As we discuss in the Appendix, our results depend weakly on the geometry of the radiation field (e.g.~beamed versus isotropic).
 We solve Eq.~(\ref{eq: H2 form_dest}) to obtain the atomic and molecular density profiles along each LOS. 
We then integrate the \ho densities to obtain the \ho column density PDF.

\subsection{Basic Parameters}
\label{sub: params and scaling}
For homogeneous gas, two parameters fully characterize the $\ho/\hh$ equilibrium problem, (i) the $\alpha G$ parameter which is proportional to $I_{\rm UV}/n$, and (ii) the dust absorption cross section $\sg$, or equivalently $\phi_g Z'$.
Since $n$ is no longer a constant when turbulent fluctuations are present, we define 
\begin{equation}
\label{eq: mean aG}
\overline{\aG} \ \equiv \ \frac{D_0 G}{R\langle n \rangle} = 2.0 \  I_{\rm UV} \left( \frac{30 \ {\rm cm^{-3}}}{\langle n \rangle} \right)  \ ,
\end{equation}
where we have replaced $n$ with the volume average $\langle n \rangle$ in Eq.~(\ref{eq: aG}).
We consider a wide range of $\overline{\aG}$ values, from the weak ($\overline{\aG} \ll 2$) to the strong  ($\overline{\aG} \gg 2$) field limits.
For $\sg$ we assume the standard value $\sg = 1.9 \times 10^{-21} \ {\rm cm^2}$ corresponding to $\phi_g Z'=1$. 
% We discuss the effects of varying metallicity in \S \ref{sec: discussion}.

The \ho column is accumulated over a typical length of
\begin{equation}
\label{eq: HI scale turbulence}
L_{\ho} \equiv \frac{1}{\sg \langle n \rangle} = 5.7 \left( \frac{30 \ {\rm cm^{-3}}}{\langle n \rangle} \right) \frac{1}{\phi_g Z'} \ {\rm pc} \ .
\end{equation} 
For a turbulent medium, the density fluctuations have typical lengths of the decorrelation scale $L_{\rm dec}$ (\S \ref{sub: LOS averages sims}). Thus, for turbulent medium the ratio   
\begin{equation} 
\label{eq: tau_dec}
\frac{L_{\rm dec}}{L_{\ho}} =  \sg \langle n \rangle L_{\rm dec} \equiv \tau_{\rm dec} \ ,
% = \ 0.2 \frac{L_{\rm drive}}{L_{\ho}} ,
\end{equation} 
enters as an additional parameter.
The $L_{\rm dec}$-to-$L_{\ho}$ ratio has the physical meaning of a mean dust opacity over the decorrelation width, denoted by $\tau_{\rm dec}$.
The ratio $L_{\rm dec}/L_{\ho}=\tau_{\rm dec}$ further determines the characteristic number of fluctuations along the \ho length, through
\begin{equation}
\label{eq: paz_N HI}
\pazocal{N}(L_{\ho}) = 1 + \frac{L_{\ho}}{L_{\rm dec}} = 1+\tau_{\rm dec}^{-1} \ ,
\end{equation} (i.e.~Eq.~\ref{eq: paz N} with ${\ell}= L_{\ho}$), which then controls the $\ho/\hh$ structure. 
Following Eqs.~(\ref{eq: L_dec L_drive},\ref{eq: tau_dec}), the driving scale is related to  $\tau_{\rm dec}$ 
through
\begin{equation}
\label{eq L_drive tau_dec}
L_{\rm drive} = 5.0 \tau_{\rm dec}L_{\ho} \ .
\end{equation}

\subsection{Profiles}
\label{sub: profiles}

We scale our simulations such that the average optical depth over the box is
$\tau_{\rm box} \equiv \sg \langle n \rangle L_{\rm box} = 5.7$ ensuring \tra conversion for all sightlines (since $\tau_{\rm box} \gg 1$).
Following Eq.~(\ref{eq: Ldec}) and (\ref{eq: paz_N HI}), $\tau_{\rm box}$ sets $\tau_{\rm dec}$ and $\pazocal{N}(L_{\ho})$, giving
$\tau_{\rm dec}=0.45$, $\pazocal{N}(L_{\ho}) = 3.2$.
We use the above scaling for our results in this section and in \S \ref{sub: HI distribution}.
Following Eqs.~(\ref{eq: t_chem t_turb ratio}) and (\ref{eq: L_dec L_drive}), $\tau_{\rm dec}=0.45$ implies $t_{\rm chem}/t_{\rm turb}=0.44 (\phi_g/\phi_R) \ms$.
Thus, for the highest Mach number we consider ($\ms=4.5$) the chemical time may exceed the turbulent time, unless the H$_2$ formation efficiency is enhanced ($\phi_R>2$), or the dust absorption efficiency  is reduced ($\phi_g<1/2$).

In the upper panels of Fig.~\ref{fig: N1_tot_profiles} we show the density profiles, $x\equiv n/\langle n \rangle$, for three arbitrary sightlines for the $\ms=0.5$, 2 and 4.5.
The cloud depth in the abscissa is represented by the mean opacity $\langle \tau({\ell}) \rangle \equiv \sg \langle n \rangle {\ell}$, ranging from 0 to $\tau_{\rm box}=5.7$.
The horizontal bars in each panel represent the decorrelation opacity width $\tau_{\rm dec}=0.45$, which is comparable to the typical scales of density fluctuations.
The middle and lower panels show the calculated \ho profiles $x_{\ho} \equiv n_{\ho}/n$, and the integrated \ho column densities, $N_{\ho}({\ell}) \equiv \int_0^{l} n_{\ho} d{\ell}' $, for the corresponding LOS, assuming $\overline{\aG}=2.0$ and $\phi_g Z'=1$.
For comparison, the dashed curves show the exponential decay of $x_{\ho}$ and the gradual  buildup of $N_{\ho}({\ell})$ as obtained by the uniform-density solution.\footnote{In our notation, $N_{\ho}({\ell})$ refers to the integrated \ho column density (that depends on cloud depth), whereas $N_{\ho} \equiv \lim_{l \rightarrow \infty} N_{\ho}(l)$ denotes the total (asymptotic) \ho column density.}
%are the corresponding profiles for homogeneous gas of the same mean density.
% The \ho and \hh profiles are the \ho and \hh fractional abundances, $x_1 \equiv n_1/n$ and $2x_2 \equiv 2n_2/n$  as functions of cloud depths, as parameterized through the accumulated hydrogen column density $N(r) = \int_0^r n(r') dr'$ (here $r$ is the coordinate along the column).

For the subsonic simulation ($\ms=0.5$), the density remains nearly homogeneous and the $\ho/\hh$ profiles and the integrated \ho column densities remain close to the homogeneous density solution.
As the Mach number increases and exceeds unity, density fluctuations become substantial, and the \ho (and $\hh$) density profiles become highly distorted.
For the highly supersonic case ($\ms=4.5$) the \ho profiles exhibit an extreme scatter, differing by orders of magnitude in some locations. For example, at cloud depth $\langle \tau({\ell}) \rangle = 3$, $x_{\ho}\approx 10^{-5}$, $2\times 10^{-2}$, and $0.8$ for the red, yellow and blue LOS.
This increasing scatter with $\ms$ is further reflected in $N_{\ho}({\ell})$ and in the total (asymptotic) \ho column density, $N_{\ho}$. 
However, because $N_{\ho}$ is an integrated quantity, the perturbations are (partially) averaged out and the scatter is much smaller than for $x_{\ho}$. 
For example, for the three LOS of $\ms=4.5$, the scatter in $N_{\ho}$ is less than 0.4 dex.  In \S \ref{sub: HI distribution} we show the calculated PDFs for the total \ho columns.

\begin{figure*}
    \includegraphics[width=1\textwidth]{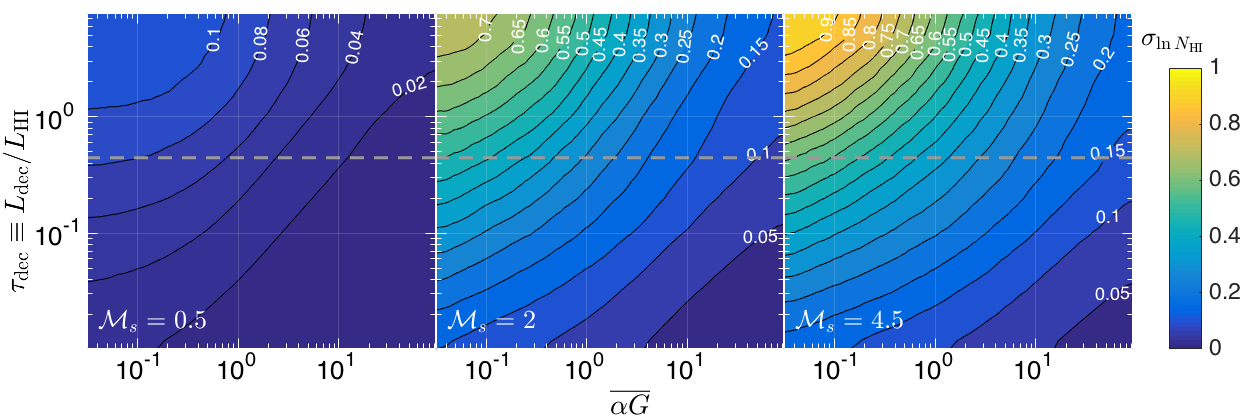}        % Turb/plots_for_paper/plot_s_lnN1_vs_Ms_aG_and_scale
	\caption{The standard deviation of $\ln N_{\ho}$ (natural logarithm) in the $\overline{\aG}-\tau_{\rm dec}$ parameter space, as calculated for the $\ms=0.5$, 2 and 4.5 simulations. The dashed line indicate the $\tau_{\rm dec}=0.45$ value used for Figs.~\ref{fig: N1_tot_profiles}-\ref{fig: N1_dists_sims}.
}	
\label{fig: slnN1_vs_aG_tau_dec}
\end{figure*}

\subsection{The \ho column density distribution}
\label{sub: HI distribution}
We integrate the \ho profiles along all of the lines-of-sight for the $\ms=0.5$, 2 and 4.5 simulations, and obtain the \ho column density distributions.
In the upper panels of Fig.~\ref{fig: N1_dists_sims} we show PDFs of $\log_{10} N_{\ho}$ for $\overline{\aG}=0.2$, 2, and 20. The lower panels show the median (solid curve), mean (dotted curve, which almost converges with the median), and the 68.3, 95.5, 99.7 percentiles (about the median) as functions of $\overline{\aG}$.
Evidently, the distributions become wide with increasing $\ms$, and with decreasing $\overline{\aG}$. 
% All simulations assume Alfv\'enic Mach number  is $M_A=0.7$, initially parallel to the $z$ direction.
% We calculate three $N_{\ho}$ distributions for three assumed directions of incident radiation, $x$, $y$ (perpendicular to $B$) and $z$ (parallel to $B$). 
% We find that the resulting $N_{\ho}$ distribution weakly depends on the assumed direction, and show the average of the three directions.
% We have also considered simulations with a weaker magnetic field, $M_A=2.0$  (not shown), and found that the resulting $N_{\ho}$ distribution is similar to the sub-Alfv\'enic case.
For example, for the $\ms=4.5$ simulation the width of the 68.3 percentile is 0.2 dex and 0.5 dex for $\overline{\aG}=10$ and 0.1 respectively.
For the $\ms=0.5$ simulation the width of the 68.3 percentile
ranges from 0.03 to 0.1 dex for $\overline{\aG}=10$ to 0.1.
The widening of the \ho PDF with increasing \ms reflects the increasing spread of the density PDF, as illustrated in Fig.~\ref{fig: maps} and in the $\sigma-\ms$ relation, Eq.~(\ref{eq: s-M relation}).
The \ho PDFs become narrow at large \aG because in the strong field limit the \tra is very sharp as the radiation is absorbed by dust (exponential attenuation), resulting in a weak dependence of the \ho column on gas density (Eq.~\ref{eq: N1tot}). 

The median and mean are above the homogeneous solution for small $\overline{\aG}$ and below it for large $\overline{\aG}$.
% The transition occurs at $\overline{\aG} \approx 2$, in the middle of the CNM range.
Importantly, for all $\overline{\aG}$, the deviations from the homogeneous solution remain small.
For example, for $\overline{\aG}$ ranging from 0.1 to 10, the deviation is 0.08 to 0.24 dex.~for the $\ms=4.5$ case, and is 0.02 to 0.13 dex for $\ms=2$.
Thus, the mean \ho column is well approximated by the \citetalias{Sternberg2014} formula for homogeneous gas, as given by Eq.~(\ref{eq: N1tot}), with $n=\langle n \rangle$.

Interestingly, the shapes of the \ho PDFs deviate from a lognormal. 
% For the supersonic simulations, the PDFs have negative skewness and exhibit a sharp cutoff at the high end. 
For all supersonic simulations, the $N_{\ho}$ distributions are strongly truncated at the highest ends, and have extended tails at the lower-end of the distribution. 
% This effect was first noticed observationally in \citet{Burkhart2015,Imara2016} with strongly truncated HI PDFs at the high column density end coincident with the HI-H$_2$ transition.
% What do you mean? In Burkhart+15 the PDF is fairly symmetric
This is due to the interaction of the propagating radiation with the non-uniform gas and the effect on the H$_2$ self-shielding.
% As discussed in \ref{sub: profiles}, 
At any point inside the cloud, the H$_2$ and \ho fractions depend on the local volume density of the gas.
However, the H$_2$ self-shielding introduces a dependence 
on the accumulated H$_2$ column, from the edge to the point of interest.
This introduces a non-linear dependence of the $\ho$/H$_2$ fractions on the volume density profile of the gas.
Any positive density perturbation along the column results in a disproportional increase of the H$_2$ fraction and reduced \ho fraction, from that point onward.
This effect introduces a bias towards lower values of $N_{\ho}$.

\subsection{$\sigma_{\ln N_{\ho}}$ versus $\overline{\aG}$, $\ms$ and $\tau_{\rm dec}$}
\label{sub: sigma_lnN1}
% As discussed in \S \ref{sub: params and scaling} the assumed simulation scale for $\tau_{\rm box}$ sets $\tau_{\rm dec}$ and $\pazocal{N}_{\ho}$ through $\tau_{\rm dec}= 6.8 \times 10^{-3} \tau_{\rm box}$, $\pazocal{N}_{\ho}=1+\tau_{\rm dec}^{-1}$.
The width of the $\ln N_{\ho}$  distribution depends on three parameters, $\overline{\aG}$, the Mach number $\ms$, and the decorrelation opacity $\tau_{\rm dec}$. 
The first encapsulates H$_2$ formation versus destruction, the second determines the width of the density distribution, and the last determines the frequency of density fluctuations along an \ho column.
% In \S \ref{sub: profiles}-\ref{sub: HI distribution} we assumed the fiducial scaling $\tau_{\rm box}=5.7$ which corresponds to $\tau_{\rm dec}=0.45$ (see \S \ref{sub: params and scaling}).
We vary $\tau_{\rm dec}$ by modifying the simulation scaling $\tau_{\rm box}$ (see Eq.~\ref{eq: Ldec}), and consider $\tau_{\rm dec}$ from 0.01 to 8.
% For small $\tau_{\rm dec}$ ($<0.45$), we attach several (randomly picked) density profiles, back-to-back, to ensure \tra transition. 

\begin{figure*}
 	\centering
	% this figure was generated using the script: Turb/N1_vs_aG and plot_HI_PDFs
            \includegraphics[width=1\textwidth]{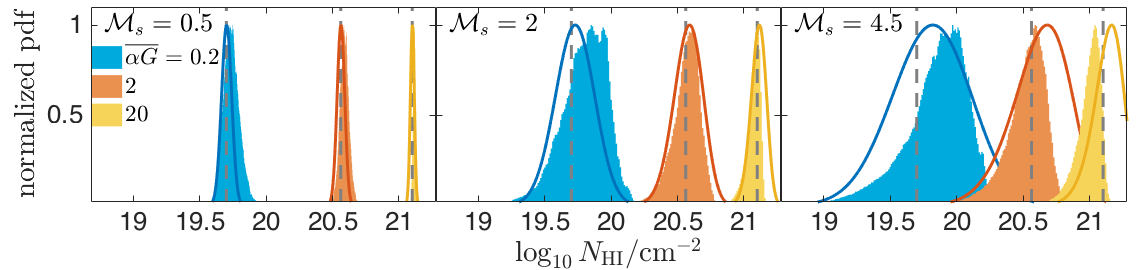}
            		\includegraphics[width=1\textwidth]{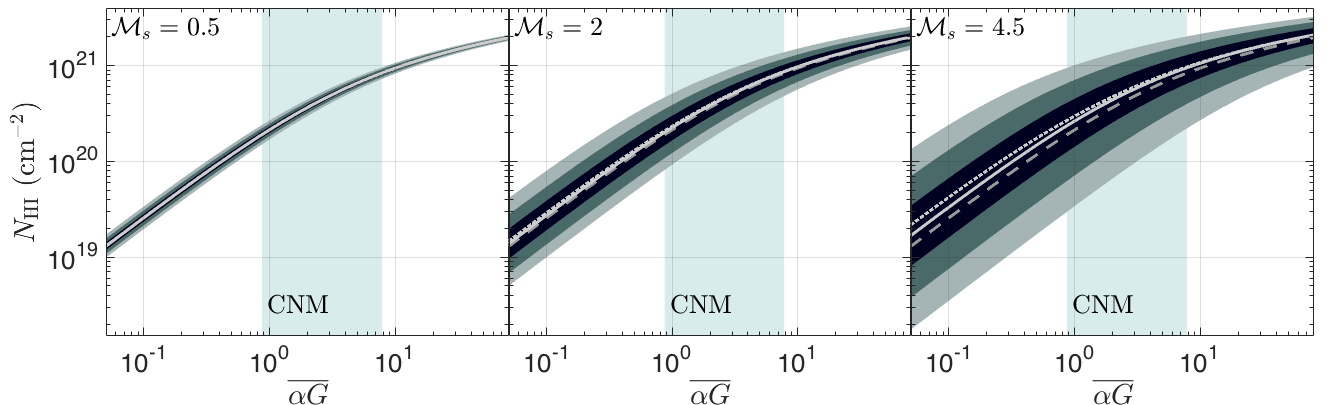}      
	\caption{Top: the PDFs of $\log_{10} N_{\ho}$, as calculated for the $\ms=0.5$, 2 and 4.5 simulations, assuming $\overline{\aG}=0.2, 2$ and 20 (shaded), and as given by Eq.~(\ref{eq: N1 dist model}) (curves). 
	Bottom: the median (solid curves), mean (dotted), and the 68.3, 95.5, 99.7 percentiles (shaded regions) as functions of $\overline{\aG}$, as given by  Eq.~(\ref{eq: N1 dist model}). 
	All panels assume $\tau_{\rm dec}=0.45$.
 The dashed curves (in all panels) are the homogeneous solutions, Eq.~(\ref{eq: N1tot}).}
    \label{fig: N1_vs_aG_sims_and_model}
\end{figure*}

% The PDFs shown in Figs.~\ref{fig: N1_dists_sims}-\ref{fig: N1_tot_vs_aG_sims} assume the fiducial scaling $\tau_{\rm box}=5.7$ corresponding to 
% we plot the standard deviation of  $\tau_{\rm dec}$
% The \ho PDFs also depend on the value of $\pazocal{N}(L_{\ho})$, and therefore on the assumed scale of the simulation.

In Fig.~\ref{fig: slnN1_vs_aG_tau_dec} we plot $\sigma_{\ln N_{\ho}}$ (natural logarithm) in the $\overline{\aG}-\tau_{\rm dec}$ plane, for the $\ms=0.5$, 2, and 4.5 simulations.
The dashed line marks the $\tau_{\rm dec}=0.45$ value used for Figs.~\ref{fig: N1_tot_profiles}-\ref{fig: N1_dists_sims}.
As discussed in \S \ref{sub: HI distribution}, $\sigma_{\ln N_{\ho}}$ increases with increasing \ms or with decreasing $\overline{\aG}$.
However, Fig.~\ref{fig: slnN1_vs_aG_tau_dec} shows that $\sigma_{\ln N_{\ho}}$ also has a strong dependence on $\tau_{\rm dec}$.
For small $\tau_{\rm dec}$, the number of density fluctuations along a LOS, $\pazocal{N}$, is large ($\pazocal{N}=1+\tau_{\rm dec}^{-1}$). Each LOS then samples a large portion of the (same) parent density PDF, and the different sightlines become more alike.
As $\tau_{\rm dec}$ increases, $\pazocal{N}$ decreases, until finally for $\tau_{\rm dec} \gg 1$, $\pazocal{N}\rightarrow 1$.
Each LOS is then correlated over the entire $L_{\ho}$ scale, and its density is approximately uniform, drawn from the parent density PDF.
The width of the \ho column density is then maximal and reflects the width of the parent density PDF.

In the following section we derive analytic formula for $\sigma_{\ln N_{\ho}}$ as a function of $\overline{\aG}$, $\ms$ and $\tau_{\rm dec}$.

\begin{figure*}
	\centering
	\includegraphics[width=1\textwidth]{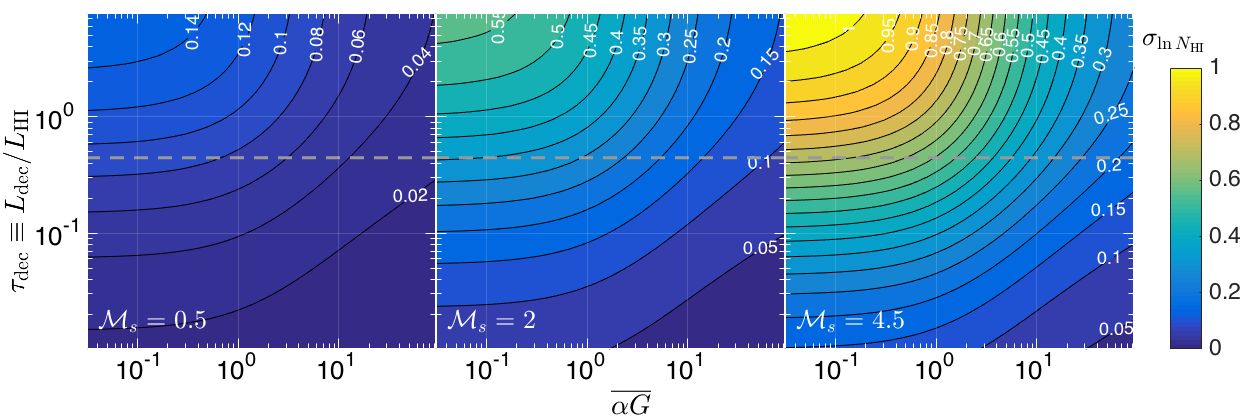}
	\caption{The standard deviation of $\ln N_{\ho}$ (natural logarithm) in the $\overline{\aG}-\tau_{\rm dec}$ parameter space, as given by Eq.~(\ref{eq: s_lnN}). The dashed line indicates $\tau_{\rm dec}=0.45$, that is used for Fig.~\ref{fig: N1_vs_aG_sims_and_model}.
		}
		\label{fig: slnN1_sims_and_model}
        %/Turb/plot_s_lnN1_vs_Ms_aG_and_scale
\end{figure*}

\section{Analytic Approximation}
\label{sec: model}
% The basic trends of the \ho column density PDFs; (a) the increasing dispersion with increasing Mach number (b) the decreasing dispersion with increasing $\overline{\aG}$, and (c) the increasing dispersion with increasing $\tau_{\rm dec}$, may be understood as follows.
As is shown in Fig.~\ref{fig: N1_tot_profiles}, each LOS has a unique density profile, with a complicated \ho density structure.
To obtain a simple analytic representation for the distribution of \ho columns, we approximate each LOS as containing a uniform density equal to the average along the \ho length. I.e.~for each LOS we set the density equal to $\langle n \rangle x_{L_{\ho}}$ (i.e.~Eq.~\ref{eq: xLdefinition} with ${\ell} = L_{\ho}$). 
Hereafter we use a shortened notation and omit the subscript \ho in $x_{L_{\ho}}$.
For each LOS, the \ho column density is then given by
\begin{equation}
\label{eq: N1tot(x)}
N_{\ho} \ = \ \frac{1}{\sg} \ \ln \left[ \frac{\overline{\aG}}{2} \frac{1}{x_L} \ + \ 1 \right] \ .
\end{equation}
The PDF of $\ln N_{\ho}$ is thus
\begin{equation}
\label{eq: N1 dist model}
\frac{\mathrm{d}f}{\mathrm{d}\ln N_{\ho}} \ = \left(1+\frac{2x_L}{\overline{\aG}} \right) \ln \left( \frac{\overline{\aG}}{2 x_L}+1 \right) \ \frac{\mathrm{d}f}{\mathrm{d} \ln x_L}  \  \ ,
\end{equation} where $\mathrm{d}f/\mathrm{d}\ln x_L$ is the PDF of $\ln x_L$.
Following the discussion in \S \ref{sub: LOS averages sims}, $x_L$ and $\ln x_L$ are approximately Gaussian and lognormal with standard deviations
\begin{align}
\label{eq: sigma xL(LHI)}
\sigma_{x_L} &=\frac{\sigma_x}{\sqrt{\pazocal{N}}} \simeq \frac{b \ms}{\sqrt{1+\tau_{\rm dec}^{-1}}} \\ 
\sigma_{\ln x_L} &=\ln^{1/2}\left( 1+\sigma_{x_L}^2 \right) \simeq \ln^{1/2}\left( 1+\frac{[b \ms]^2}{1+\tau_{\rm dec}^{-1}} \right) \ ,
\label{eq: sigma ln_xL}
% \left(1+\frac{1}{L_{\rm dec}\langle n \rangle \sg} \right)^{-1/2} \ , 
\end{align} respectively. 
where the second equality follows from Eqs.~(\ref{eq: s-M relation}) and (\ref{eq: paz_N HI}).

Fig.~\ref{fig: N1_vs_aG_sims_and_model} shows the $\ln N_{\ho}$ PDFs as obtained from the simulations (also shown in Fig.~\ref{fig: N1_vs_aG_sims_and_model}), along with the analytic PDFs, as given by Eq.~(\ref{eq: N1 dist model}).
The locations and the widths of the analytic PDFs roughly follow the PDFs from the simulations, with some  differences.
First, the shapes of the analytic PDFs are symmetric whereas the simulated PDFs are truncated at the high end and have left-tails.
This difference is expected because the analytic approximation does not account for the interaction of the radiation with the density perturbations along each LOS, and as discussed in \S \ref{sub: HI distribution}, this interaction introduces a preference for small \ho columns. 
%An additional deviation is caused because the analytic PDFs assume the $\sigma_x-\ms$ relation which is only approximately satisfied in the simulations (see the lower panel in Fig.~\ref{fig: maps}). 

The lower panels of Fig.~\ref{fig: N1_vs_aG_sims_and_model} show the median (solid), mean (dotted), and the 68.3, 95.5, 99.7 percentiles about the median (shaded regions), as functions of $\overline{\aG}$. 
The dashed curves are the homogeneous solutions, for comparison.
The trend of an increasing dispersion with increasing \ms or with decreasing $\overline{\aG}$ is in agreement with the numerical results shown in Fig.~\ref{fig: N1_dists_sims}. 
The median and mean \ho columns remain close to the homogeneous solution, also in agreement with the numerical results.
An analytic expression for the median $N_{\ho}$ is obtained by substituting the median
\begin{equation}
\label{eq: x_med}
x_{L,{\rm med}} = \mathrm{e}^{-\frac{1}{2}\sigma_{\ln x_L}} \simeq \left( 1+\frac{[b\ms]^2}{1+\tau_{\rm dec}^{-1}} \right)^{-1/2} ,
\end{equation} into Eq.~(\ref{eq: N1tot(x)}), giving
\begin{equation}
\label{eq: x_med}
N_{\ho,{\rm med}} = \frac{1}{\sg} \ln \left[ \frac{\overline{\aG}}{2}\Big(1+ \frac{[b\ms]^2}{1+\tau_{\rm dec}^{-1}} \Big)^{1/2} +1 \right] \ .
\end{equation}
For moderate $b\ms$ (or if $\tau_{\rm dec} \ll 1$), $x_{L,{\rm med}} \approx 1$ and the median $N_{\ho}$ remain close to the homogeneous solution (see also Fig.~\ref{fig: N1_vs_aG_sims_and_model}).
% For example, for $b=1/3$, $\ms=4.5$, and $\tau_{\rm dec} = 0.1, 0.5$, and 1, $x_{L,{\rm med}}=0.9$, 0.8, and 0.7, and $N_{\ho,{\rm med}}$ deviates from the homogeneous 
Then, to a good approximation
\begin{align}
\label{eq: N_med N_mean}
\langle N_{\ho} \rangle \simeq N_{\rm \ho,med}  \simeq \frac{1}{\sg} \ln \left[ \frac{\overline{\aG}}{2}+1 \right]  \ ,
% \\  \nonumber
% &=  5.3 \times 10^{20} \frac{1}{\phi Z'} \ln \left[ \frac{\overline{\aG}}{2}+1 \right] \ {\rm cm^{-2}}
\end{align}
similar to Eq.~(\ref{eq: N1tot}) for homogeneous gas, but with $\aG$ replaced by $\overline{\aG}$.
This result is also confirmed by our numerical computations shown in Fig.~\ref{fig: N1_dists_sims}.

The standard deviation of $\ln N_{\ho}$ may be approximated by
\begin{equation}
\sigma_{\ln N_{\ho}} \simeq \sigma_{\ln x_L} \frac{\mathrm{d}\ln N_{\ho}}{\mathrm{d}\ln x_L} \Big|_{\ln x_L=0} \ ,
\end{equation}
where the approximation becomes increasingly accurate for small $\sigma_{\ln N_{\ho}}$ values. Plugging in  Eq.~(\ref{eq: N1tot(x)}) and (\ref{eq: sigma ln_xL}) we get
\begin{align}	
\label{eq: s_lnN}
% \label{eq: s_lnN with NHI}
% \sigma_{\ln N_{\ho}} & \simeq \sigma_{\ln xl} \left( 1+\frac{2}{\overline{
% \aG}} \right)^{-1} \frac{1}{\sg \langle N_{\ho} \rangle} 
\sigma_{\ln N_{\ho}} \simeq  \ddfrac{\ln^{1/2}\left(1+\frac{[b\ms]^2}{1+\tau_{\rm dec}^{-1}} \right)}{\left( 1+\frac{2}{\overline{\aG}} \right) \ln \left( \frac{\overline{\aG}}{2}+1 \right)  } \ .
\end{align}
% The standard deviation of $N_{\ho}/\langle N_{\ho} \rangle$ is related to $\sigma_{\ln N_{\ho}}$ through $\sigma_{\ln N_{\ho}}^2 \simeq \ln (1+\sigma_{N_{\ho}/\langle N_{\ho} \rangle}^2)$, and for small dispersions, $\sigma_{\ln N_{\ho}} \simeq \sigma_{N_{\ho}/\langle N_{\ho} \rangle} =  \sigma_{N_{\ho}}/\langle N_{\ho} \rangle$.
In this expression, the nominator is  $\sigma_{\ln x_L}$ (Eq.~\ref{eq: sigma ln_xL}), introducing the dependence on the turbulence parameters, $b$, \ms and $\tau_{\rm dec}$. 
As expected, $\sigma_{\ln N_{\ho}}$ increases with increasing $b \ms$.
$\sigma_{\ln N_{\ho}}$ increases with $\tau_{\rm dec}$, and becomes independent of $\tau_{\rm dec}$ once $\tau_{\rm dec} \gg 1$.
For $\tau_{\rm dec} \rightarrow 0$, $\sigma_{\ln N_{\ho}} \propto \tau_{\rm dec} \rightarrow 0$.
The dependence on the radiation intensity enters  through the $\overline{\aG}$ parameter (Eq.~\ref{eq: mean aG}).
For $\overline{\aG} \ll 1$ (the weak field limit), $\sigma_{\ln N_{\ho}} \simeq \sigma_{\ln x_L}$ and the dispersion is maximal and is independent of $\overline{\aG}$.
For $\overline{\aG} \gg 1$ (the strong field limit), $\sigma_{\ln N_{\ho}} \simeq \sigma_{\ln x_L}/\ln(\overline{\aG}/2)$, and the distribution becomes narrow with increasing $\overline{\aG}$.

Fig.~\ref{fig: slnN1_sims_and_model} shows $\sigma_{\ln N_{\ho}}$ as a function of $\tau_{\rm dec}$ and $\overline{\aG}$, as given by Eq.~(\ref{eq: s_lnN}).
Like the numerical results (shown in Fig.~\ref{fig: slnN1_vs_aG_tau_dec}), the standard deviation increases as (a)  $\ms$ increases, (b) as $\tau_{\rm dec}$ increases, and (c) as $\overline{\aG}$ decreases.
Deviations from the numerical results exist, and are expected given that the analytic model introduces simplifying assumptions, (a) the density correlations are described by a single decorrelation scale, $\tau_{\rm dec}$, (b)  the density distribution is lognormal and follows the $\sigma_x - \ms$ relation (Eq.~\ref{eq: s-M relation}), and (c) the ansatz that for each LOS the \ho column is given by Eq.~(\ref{eq: N1tot(x)}). 
The advantage of the analytic approximation is that it provides a smooth solution for $\sigma_{\ln N_{\ho}}$ as a function of $\ms$ and $b$.

\section{Applications to Observations}
\label{sec: observations}

In this section we present a brief example demonstrating how our results for the width and mean of the \ho column PDF may be used to analyze 21 cm observations toward molecular clouds, setting constrains on the Mach number and turbulence driving scale.

Based on 21 cm emission lines from the GALFA-$\ho$ Survey \citep{Peek2011}, \citet{Lee2012} obtained an \ho map for the Perseus molecular cloud.
\citet{Burkhart2015} derived the \ho PDF and  found that it to be very narrow, with $\sigma_{N_{\ho}}/\langle N_{\ho}\rangle = 0.13$.
Based on absorption line data from \citet{Stanimirovic2014}, \citet{Burkhart2015} 
obtained the Mach number distribution for the cold neutral medium (CNM) around Perseus. They find that $\ms$ ranges from $\ms=1$ to 11, with a median value $\ms=4$.
 
Unlike the Mach number, the \ho PDF, being observed in emission, contains contributions from both the WNM and the CNM phases.
The former being typically subsonic, and the latter supersonic \citep{Heiles2003,Wolfire2003}.
We decompose the observed $\sigma_{N_{\ho}/\langle N_{\ho}\rangle}  = 0.13$ into CNM and WNM components,
\begin{equation}
\label{eq: s CNM-WNM}
\sigma_{N_{\ho}/\langle N_{\ho}\rangle} = \left(\phi_C \sigma_{N_{\ho, C/\langle N_{\ho,C}\rangle}}^2 + \phi_W \sigma_{N_{\ho, W/\langle N_{\ho,W}\rangle}}^2 \right)^{1/2}  ,
\end{equation}
where the subscripts $C$ and $W$ refer to CNM and WNM, and $\phi_C$ and $\phi_W$ are the gas mass fractions in these phases.
\citet{Stanimirovic2014} obtained that around Perseus, $\phi_C$ ranges between 0.1 and 0.5, with a median $\phi_{C}=0.35$. Assuming the median $\phi_{C}=0.35$, and assuming that the WNM is subsonic and thus has a negligible \ho dispersion (see Figs.~\ref{fig: N1_dists_sims}-\ref{fig: slnN1_sims_and_model} above), we obtain
$\sigma_{N_{\ho, C/\langle N_{\ho,C}\rangle}}=0.22$, or equivalently\footnote{For a lognormal distribution, $\sigma_{\ln N_{\ho}} = \ln^{1/2}(1+\sigma_{N_{\ho}/\langle N_{\ho}\rangle}^2)$.} $\sigma_{\ln N_{\ho, C}}=0.22$, for  the CNM in Perseus.

Inspecting the $\ms=4.5$ panel in Fig.~\ref{fig: slnN1_vs_aG_tau_dec}, we see that for $\overline{\aG}$ within the CNM range $(\overline{\aG})_{\rm CNM}=1-8$ (see \S \ref{sec: Theory}), the $0.22$ contour is obtained for $\tau_{\rm dec}=L_{\rm dec}/L_{\ho}=0.06$ to $0.3$, respectively.
These values of $\tau_{\rm dec}$ correspond to $L_{\rm drive}/L_{\ho}=0.3-1.5$ (Eq.~\ref{eq L_drive tau_dec}), i.e.~the driving scale is of order of the \ho scale-length.
Assuming typical CNM density, $\langle n \rangle \approx 30$~cm$^{-3}$, and for $\phi_g Z'=1-2$ as suggested by \citet{Lee2012}  for Perseus, we obtain (with Eq.~\ref{eq: HI scale turbulence}) $L_{\rm drive} \sim 1-8$~pc.

If we do not neglect the WNM contribution in Eq.~(\ref{eq: s CNM-WNM}), $\sigma_{\ln N_{\ho,C}}$ would be smaller, further reducing $\tau_{\rm dec}$ and $L_{\rm drive}$.
Our numerical computations in Fig.~\ref{fig: slnN1_vs_aG_tau_dec} are based on the MHD simulations that assume solenoidal driving ($b=1/3$). For compressional driving or mixed driving the contours in Fig.~\ref{fig: slnN1_vs_aG_tau_dec} would shift downwards (since increasing $b$ is similar to increasing $\ms$; Eq.~\ref{eq: s-M relation}) and $\tau_{\rm dec}$ and $L_{\rm drive}$ will again decrease.
Thus, our derived driving scale is an upper limit.

\begin{figure}[t]
	\centering	\includegraphics[width=0.5\textwidth]{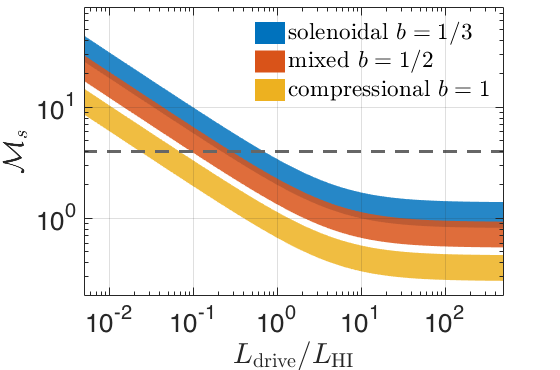}
    %generated with /Turb/plots_for_paper/plot_Perseus
	\caption{
    The relation between the CNM Mach number, $\ms$, and the driving-to-H{\scriptsize I} scale ratio, $L_{\rm drive}/L_{\ho}$, assuming different driving $b$ parameters (Eq.~\ref{eq: b_MS perseus}), as constrained from 21 cm observations towards the Perseus molecular cloud. 
The width of each strip corresponds to the $(\overline{\aG})_{\rm CNM}=1-8$ range. 
The dashed line is the median \ms obtained from absorption line data.    }
		\label{fig: Perseus}
\end{figure}

A similar conclusion may be drawn from the analytic model.
Eq.~(\ref{eq: s_lnN}) predicts the standard deviation of $\ln N_{\ho}$ as a function of $b$, $\ms$, $\tau_{\rm dec}$ and $\overline{\aG}$.
Plugging in $\sigma_{\ln N_{\ho,C}}=0.22$ and $\overline{\aG}=1-8$, and inverting Eq.~(\ref{eq: s_lnN}) we get
\begin{equation}
\label{eq: b_MS perseus}
b \ms = C \sqrt{1+\tau_{\rm dec}^{-1}}
\end{equation}
where $C=0.28-0.47$ for $\overline{\aG}=1-8$ respectively.
This relation between the Mach number, the driving forcing parameter $b$ and the driving scale ($\tau_{\rm dec}=0.2L_{\rm drive}/L_{\ho}$; Eq.~\ref{eq L_drive tau_dec}) is shown in Fig.~\ref{fig: Perseus} for $b=1/3$, 1/2 and 1, corresponding to solenoidal, mixed and compressional drivings.
The width of the strips correspond to the width of the $\overline{\aG}=1-8$ CNM range.
For the median $\ms=4$ (dashed line), $L_{\rm drive}/L_{\ho}$ ranges within 0.2-0.7, 0.1-0.3, and 0.02-0.07 for $b=1/3$, $b=1/2$ and $b=1$.

Interestingly, narrow \ho PDFs were recently reported for more Galactic clouds: Ophiuchus, Orion A, Orion B, California, MonR2 and Rosette \citep{Imara2016}, suggesting that the presence of a small
driving scale, of order of, or smaller than the \ho length, might be a general feature in molecular clouds and/or a considerable WNM component to the HI column density.

However, there are caveats to this analysis.
First, the decomposition of the \ho PDF from emission line measurements into CNM and WNM components, which requires CNM/WNM fraction from absorption measurements, is uncertain. 
Thus, it would be valuable to use absorption line measurements directly to infer the \ho PDF of the CNM.
Second, accurate measurements of the Mach number are  difficult due to line of sight blending of features in position-position-velocity space.
Third, the width of the \ho PDF may be also affected by (a) variations in $I_{\rm UV}$ across the observed region, (b) the chosen cutoff criteria of the observed \ho cube; the velocity range, and the spatial extent, (c) the finite angular resolution which may smooth out density fluctuations, and (d) optically thick $\ho$ that may produce artificially narrow \ho PDFs \citep{Burkhart2013b}.
While for Perseus optical thickness is probably not a major issue (\citealt{Lee2015} obtain $\sim 10 \%$ corrections for the optically thick gas), it might be important in other GMCs.
A thorough observational analysis that addresses these uncertainties will be provided elsewhere. 
Model limitations are discussed in \S \ref{sec: discussion}.

\section{Summary and Conclusions}
\label{sec: discussion}
In this paper we have studied the \tra transition, and the $\ho$ column densities, maintained by far-UV radiation, in turbulent  media. 
We have used a suite of MHD simulations to produce realistic turbulent density distributions (\S \ref{sec: sims}).
The density field is nearly lognormal and the dispersion follows the PDF variance-sonic Mach number ($\sigma-\ms$) relation (\S \ref{sec: sims}). We find that for supersonic gas, the density decorrelation length is related to the driving scale through $L_{\rm dec}=0.2 L_{\rm drive}$. 
The decorrelation length determines the number of density fluctuations along a LOS, through $\pazocal{N}=1+{\ell}/L_{\rm dec}$, where $\ell$ is the length-scale along the LOS.
This simple description, allows us to model the scale dependent LOS averaged density distribution (\S \ref{sub: LOS averages sims}) and the \ho column density PDF analytically.
We note that while pure MHD simulations are scale-free, when the depth dependent \tra transition is of interest (or any other chemical species), the adopted simulation box length plays an important role. 
The adopted box scale determines $L_{\rm dec}$, which in turn defines the number of density fluctuations along an \ho column ($\pazocal{N}=1+L_{\ho}/L_{\rm dec}$), thus affecting the $\ho/\hh$ structure
(see \S \ref{sub: params and scaling} and \S \ref{sub: sigma_lnN1}).

As we show in \S \ref{sub: profiles}, once the turbulence becomes supersonic, strong density fluctuations are developed, and the atomic-to-molecular density profiles are significantly distorted relative to those for homogeneous uniform density media.
As a result, different lines of sight (LOS) differ in their total accumulated \ho column density, $N_{\ho}$. 
We calculate the probability density function (PDF) for $N_{\ho}$ (\S \ref{sub: HI distribution}-\ref{sub: sigma_lnN1}), as a function of the governing physical parameters: (a) the sonic Mach number $\ms$, (b) the effective dissociation parameter $\overline{\aG}$ and (c) the decorrelation opacity $\tau_{\rm dec}$, which is related to the turbulence driving scale  ($\tau_{\rm dec} \equiv L_{\rm dec}/L_{\ho} = 0.2 L_{\rm drive}/L_{\ho}$). 

We find that the mean and median $N_{\ho}$ are affected by turbulence, but as long as $b \ms \sim 1$, or if $\tau_{\rm dec} \ll 1$, $N_{\ho}$ may be well approximated by the \citetalias{Sternberg2014} and \citetalias{Bialy2016} formula for uniform-density gas, 
\begin{equation*}
N_{\ho}=\frac{1}{\sg} \ln\left(\frac{\overline{\aG}}{2}+1\right) \ \ \ ,
\end{equation*}
 where
\begin{equation*}
\overline{\aG} \equiv 2.0 \  I_{\rm UV} \left( \frac{30 \ {\rm cm^{-3}}}{\langle n \rangle} \right) \ 
\end{equation*} 
is the effective dimensionless dissociation parameter, where $I_{\rm UV}$ is the free-space intensity of the far-UV field and $\langle n \rangle$ is the volume density. Here \sg is (cm$^{2}$) is the dust-grain absorption cross section per hydrogen nuclei averaged over the Lyman-Werner dissociation band. 

The major effect of turbulent density fluctuations, is in producing a spread in the \ho column distribution.
For subsonic gas the density is nearly uniform and the \ho PDF is very narrow and the solutions converge to the \citetalias{Sternberg2014} and \citetalias{Bialy2016} formula for uniform-density gas. 
As the Mach number increases, density fluctuations becomes substantial and the \ho PDF also widens.
As discussed in \S \ref{sec: H-H2 turb gas}, the \ho PDF also depends on $\overline{\aG}$ and $\tau_{\rm dec}$, becoming wider for small $\overline{\aG}$ or small $\tau_{\rm dec}$. 
In \S \ref{sec: model}, we present an analytic formula, Eq.~(\ref{eq: s_lnN}),  for the standard deviation of the \ho PDF as a function $\ms$, $\overline{\aG}$, and $\tau_{\rm dec}$.

We demonstrate how our model may be combined with 21 cm observations toward GMCs to constrain turbulent parameters (\S \ref{sec: observations}).
For Perseus (and in other Galactic clouds), the very narrow observed \ho PDF may suggest small-scale driving, potentially pointing to the importance of multi-scale turbulent driving in the CNM \citep{Heverkorn2008,Yoo2014}.
Alternatively, the narrow PDF may be caused by small-scale decorrelation lengths  induced by the abrupt change in the chemical and thermal properties at the \tra transition.
Observational caveats are discussed in \S \ref{sec: observations}.

Our results are for single-phased gas, irradiated by far-UV radiation. In \S \ref{sec: observations} we describe how our results may be still applied to a mixed CNM/WNM medium, given estimates of the CNM and WNM fractions.
This marks the importance of observations of both emission and absorption 21 cm lines, which provide constrains on the CNM fraction and its Mach number.

The values of observed \ho column densities may depend on geometry (i.e.~slabs versus spheres, or slab inclination) and the number of \tra transition layers along the observed sightlines. 
However, importantly, the turbulent parameters, $\ms$ and $\tau_{\rm dec}$ are obtained from the standard deviation of the logarithmic \ho column, $\sigma_{\ln N_{\ho}}$ (Eq.~\ref{eq: s_lnN}). Thus, our analysis is robust against 
any multiplication of the \ho column, including, number of clouds, geometry factors, and inclination corrections.

A basic assumption made in our analysis is that the \ho and \hh are in chemical steady state (see \S \ref{sec: den flucs} for a discussion of the timescales). 
The coupling of time dependent chemistry and turbulence may affect the \ho PDF in various ways.
For example, rapidly changing density fluctuations may alter the dispersion in the \ho distribution, since \ho and \hh in different LOS do not have time to react to the fast density changes.
Turbulent mixing, may increase the mean \ho column by transferring molecular gas from inner shielded to outer unshielded regions, where it can rapidly dissociate and form $\ho$.
% This process is expected to not be compensated by the opposite mixing of \ho being transferred to inner shielded regions where it forms $\hh$, because \hh formation is orders of magnitude slower than \ho photodissociation.  
We plan to investigate time-dependent effects in a future work.

% In our paper we followed a twofold approach, (a) numerical simulation accompanied by \ho-\hh calculation, and (b) an analytic approximation.
% The use of both methods is important as each has its own individual limitations and advantages.  For example,  our analytic model ignores the interaction of radiation with the density perturbaions along the columns, which we do take into account in the simulations.
% Meanwhile current numerical simulations are limited in terms of having the necessary resolution to properly resolve the relationship between the density PDF dispersion and the sonic Mach number (Eq.~\ref{eq: s-M relation}),  \citep{Price2011}. In addition, the analytic model provides clear relations

% A further complication arises
% % to observational tests in emission is 
% since our model includes an single isothermal CNM distribution of density, while the WNM will introduce additional thermal fluctuations. 
% To account for these complications, further tests of our analytic model can be conducted with numerical simulations that include a two phase medium (e.g. \citet{Saury2014}) to study the impact of radiation field, turbulence, and the presence of the WNM on the observable \ho column density PDF.

We conclude that the atomic-to-molecular density profiles and the \ho column density are affected by the turbulent nature of the CNM.
For moderate Mach numbers, the homogeneous solution still provides a good estimate for the mean \ho column density.
The standard deviation of the \ho PDF contains a useful information regarding the turbulence properties.
Our model, combined with 21 cm observations, may be used to constrain the sonic Mach number and turbulence driving scale of cold atomic gas.

\acknowledgments

We thank Enrique Vazquez-Semadeni,  Sahar Shahaf and Ewine van Dishoeck for fruitful discussions.
We thank the referee for constructive comments that improved our paper.
This work was supported in part by the PBC Israel Science Foundation I-CORE Program grant 1829/12, by DFG/DIP grant STE1869/2-1 GE625/17-1, and by the Raymond and Beverly Sackler Tel Aviv University - Harvard/ITC Astronomy Program. 
B.B. acknowledges support from the NASA Einstein Postdoctoral Fellowship and the Raymond and Beverly Sackler TAU-ITC Visiting Researcher fund.

\appendix
{\renewcommand{\theequation}{A\arabic{equation}}
\section{Beamed versus isotropic irradiation}

In our radiative transfer analysis we have assumed unidirectional beamed irradiation.
Beamed irradiation is expected in the proximity of strong far-UV sources.
For clouds immersed in the diffuse Galactic  interstellar radiation field, isotropic irradiation may be a better approximation.
\citetalias{Sternberg2014} showed that for uniform-density clouds, exposed to isotropic fields the \ho column density is given by
\begin{equation}
\label{eq: NHI isotropic}
N_{\ho} \ = \ \frac{\langle \mu \rangle}{\sg} \ln \left[ \frac{\aG}{4 \langle \mu \rangle} + 1 \right] \ , 
\end{equation} where $\langle \mu \rangle \equiv 0.8$ is a geometrical factor.
For a given $\aG$, the flux (normal to the slab) for isotropic radiation is half the beamed flux, and hence the \ho column density is smaller (see \S 2.3 in \citetalias{Sternberg2014} for a full discussion and derivation of Eq.~\ref{eq: NHI isotropic}). 
Comparing Eq.~(\ref{eq: NHI isotropic}) with the beamed solution (Eq.~{\ref{eq: N1tot}}), we see that
\begin{equation}
\label{eq:iso-beamedtrans}
N_{\ho}^{\rm iso}(\aG) \ = \ 0.8 \ N_{\ho}^{\rm bm}(\aG/1.6) \ ,
\end{equation} where  $N_{\ho}^{\rm iso}$ and $N_{\ho}^{\rm bm}$ are the \ho columns produced by isotropic and beamed fields, respectively.
Eq.~(\ref{eq:iso-beamedtrans}) provides a simple transformation between the beamed and the isotropic solutions, for uniform-density gas. 
Such a transformation is particularly useful given that an isotropic field calculation is much more time consuming.

However, Eq.~(\ref{eq:iso-beamedtrans}) may be less accurate for turbulent gas,  because the density fluctuations will have different effects on the $\ho/\hh$ structures, for the different field geometries.
As an illustrative example, consider a diffuse medium, and a very dense clump. 
Let the clump be a thin disk of radius $R$, located near the slab surface at $z=r=0$ (cylindrical coordinates), where $z$ is the axis normal to the slab along which we integrate $N_{\ho}$. 
For beamed radiation, all rays originating at $r\leq R$ will pass through the clump and $N_{\ho}$ would be very small for these sightlines, whereas for LOS with $r > R$, the gas is diffuse and $N_{\ho}$ would be very large.
% The obtained $N_{\ho}$ map would have a sharp step at $r=R$.
For isotropic radiation, inclined rays (originating at $r>R$) would  penetrate behind the clump and thus $N_{\ho}(r \leq R)$ would increase compared to the beamed case.
On the other hand, for the isotropic field, gas at $r>R$ may be still partially shielded by the clump leading to a {\it decrease} in $N_{\ho}(r > R)$ compared to the beamed case. 
These two opposite effects will tend to smooth out the contrast in the $N_{\ho}$ map, and also reduce the width of the \ho PDF.

To test the effects of isotropic versus beamed irradiation, we have carried out a calculation of the $\ho/\hh$ structure for isotropic radiation impinging the $\ms=4.5$ simulation box, with $\overline{\aG}=2.6$ (the mean CNM value).
The resulting \ho column density map is shown in Fig.~\ref{fig: iso_vs_beamed} (left panel).
The corresponding PDF is shown in the right panel (red shaded).
For comparison, the middle panel, and the gray histogram in the right panel, show $0.8 \times N_{\ho}$ for beamed irradiation with $\overline{\aG}=2.6/1.6=1.63$ (as suggested by the scaling of \ref{eq:iso-beamedtrans}).
We see that the two maps and PDFs are not identical, as would be the case for a uniform-density medium.
For the isotropic case, the \ho map is smoother and the low-$N_{\ho}$ features are more extended, as expected.
The width of the isotropic PDF is somewhat narrower ($\sigma_{\ln N_{\ho}}=0.15$), and has a lower mean ($\langle N_{\ho} \rangle=1.5 \times 10^{20}$~cm$^{-2}$) compared to the beamed field ($\sigma_{\ln N_{\ho}} =0.18$, $N_{\ho}=2.1 \times 10^{20}$~cm$^{-2}$).

\begin{figure*}[t]
	\centering
	\includegraphics[width=1\textwidth]{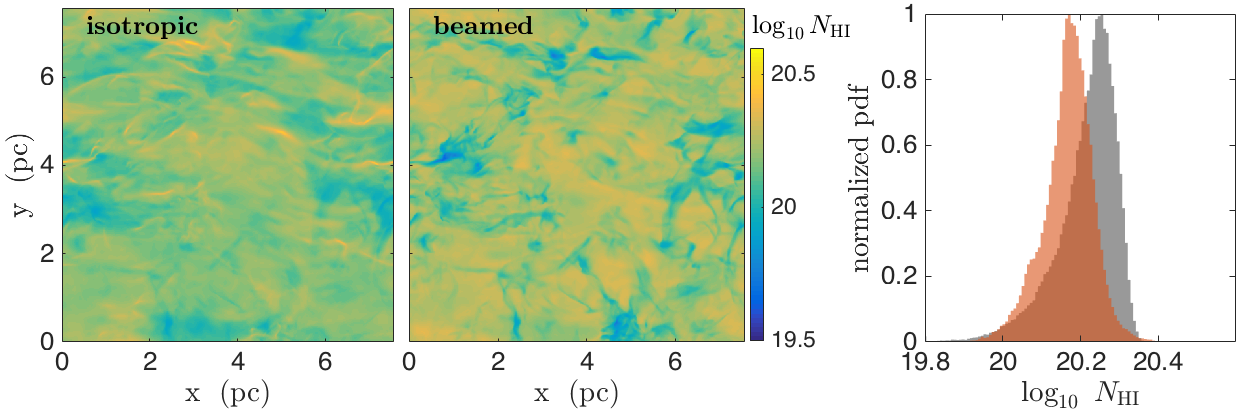} %/Turb//plots_for_paper/profiles_and_N1_tot_vs_b
	\caption{
    Left: The H{\scriptsize I} column density ($N_{\ho}$) map as calculated for the $\ms=4.5$ simulation assuming an isotropic irradiation, $\overline{\aG}=2.6$, and $\tau_{\rm dec} =0.15$. 
Middle: a map of $0.8 \times N_{\ho}$ obtained for beamed irradiation with $\overline{\aG}=1.63$, as suggested by the transformation of Eq.~(\ref{eq:iso-beamedtrans}). Right: the corresponding PDFs for the two maps (red=isotropic, gray=beamed).
    While $N_{\ho}$ may vary significantly for individual sightlines, the differences in the statistical properties of the PDFs are small.
		}
		\label{fig: iso_vs_beamed}
\end{figure*}

Importantly, while there are differences between the beamed and isotropic cases, especially when considering individual lines of sight, the relative differences in the mean and standard deviations are small ($28 \%$ and $18 \%$, respectively). 
We conclude that to a good approximation, Eq.~(\ref{eq:iso-beamedtrans}) may be used as a transformation from our (efficiently computed) beamed field results to isotropic radiation fields.

}
\bibliography{Turb}

\end{document}